\documentclass[aip,preprint]{revtex4-1}

\usepackage{amsmath,amsmath,mathrsfs,graphicx}
\usepackage{color,fancyhdr,epsf,psfrag,lineno,hyperref,float,subfigure}
\modulolinenumbers[5]
\graphicspath{ {figs/} }
\usepackage{bm}

\definecolor{R1}{rgb}{0.0,0.0,0.0} 
\definecolor{R3}{rgb}{0.0,0.0,0.0} 
\definecolor{R2}{rgb}{0.0,0,0.0} 

\draft 

\begin{document}

\title{Direct numerical simulations of the Taylor-Green Vortex interacting with a hydrogen diffusion flame: Reynolds number and non-unity Lewis number effects}

\author{Yifan Xu}
    \affiliation{State Key Laboratory of Turbulence and Complex Systems, Aeronautics and Astronautics, College of Engineering, Peking University, Beijing 100871, China}	\affiliation{AI for Science Institute, Beijing, 100080, China}

\author{Zhi X. Chen}
	\email[Author to whom correspondence should be addressed: ]{chenzhi@pku.edu.cn (Z.X. Chen).}
	\affiliation{State Key Laboratory of Turbulence and Complex Systems, Aeronautics and Astronautics, College of Engineering, Peking University, Beijing 100871, China}
	\affiliation{AI for Science Institute, Beijing, 100080, China}	

\date{\today}

\begin{abstract}
Understanding the interactions between hydrogen flame and turbulent vortices is important for developing the next-generation carbon neutral combustion systems. In the present work, we perform several direct numerical simulation (DNS) cases to study the dynamics of a hydrogen diffusion flame embedded in the Taylor-Green Vortex (TGV). The evolution of flame and vortex is investigated for a range of initial Reynolds numbers up to 3200 with different mass diffusion models. We show that the vortices dissipate rapidly in cases at low Reynolds numbers, while the consistent stretching, splitting and twisting of vortex tubes are observed in cases with evident turbulence transition at high Reynolds numbers. Regarding the interactions between the flame and vortex, it is demonstrated that the heat release generated by the flame has suppression effects on the turbulence intensity and its development of the TGV. Meanwhile, the intense turbulence provides abundant kinetic energy, accelerating the mixing of the diffusion flame with a contribution to higher strain rate and larger curvatures of the flame. Considering the effects of non-unity Lewis number, it is revealed that the flame strength is more intense in the cases with mixture averaged model. However, this effect is relatively suppressed under the impacts of the intense turbulence. 
\end{abstract}


\maketitle

\section{Introduction}
Interactions between diffusion flame and vortices are of great importance in many practical scenarios where the coupling between fluid dynamics and combustion is essential\cite{renard_dynamics_2000}. Generally, vortices make contributions to the effects of curvature and unsteadiness of diffusion flame\cite{cuenot_effects_1994}, while the heat release generated by combustion would in turn influence the enstrophy transport in turbulent flow\cite{kazbekov_enstrophy_2019,fillo_assessing_nodate}. Furthermore, quenching processes occur when the flame is submitted to external perturbations such as heat losses or strong stretch\cite{poinsot_quenching_1991,roberts_images_1993}. Such interactions drive a large class of combustion instabilities~\cite{hermanns_dynamics_2007}, which are undesirable in practical combustion devices. 

As a promising technique to decarbonising the existing gas turbine engines, direct injection (or addition in a non-premixed way) of hydrogen has gained much attention~\cite{AGOSTINELLI2022112120,ANIELLO2023112595}. Thus, understanding the interaction between hydrogen diffusion flame and vortical structures in turbulent flow is necessary for such techniques to be applied in next-generation gas turbines. 
Direct numerical simulation (DNS) plays a crucial rule in understanding the fundamental turbulent combustion processes because of its capability to not only fully resolve the turbulent flow field but also verify the validity of various models~\cite{poinsot_theoretical_2005}. Unlike the premixed flame-vortex interaction, which has been extensively studied in homogeneous isotropic turbulence, mixing shearing layer or jet configurations~\cite{DOMINGO2022}, diffusion or non-premixed flames have received much less attention and the underlying physics of their interaction with vortex requires further investigation. 

The Taylor-Green Vortex (TGV) is a well-defined and widely used configuration to study vortex evolution and turbulence transition\cite{TaylorGreen}. It has also been considered as a challenging benchmark for the validation of accuracy and efficiency of Computational Fluid Dynamics (CFD) codes\cite{WangTGV}. Numerous studies on the TGV have already been performed in the turbulence research\cite{sharma_vorticity_2019,brachet_taylor-green_1984,lusher_assessment_2021,peng_effects_2018}. However, it is seldom used in the area of combustion. Zhou et al.\cite{yang_evolution_2011}first combined the TGV configuration with a hydrogen/air premixed flame to explore the evolution of the flame front and vortex surfaces based on vortex-surface field (VSF) method. Abdelsamie et al.\cite{abdelsamie_taylorgreen_2021} combined the TGV configuration with a hydrogen diffusion flame to develop a suitable benchmark for reacting flows simulated using high-order codes. While this highly standardised configuration has been seen as very useful in many aspects (e.g. verifying numerical accuracy, testing computational efficiency, profiling parallel scalability) for newly developed reacting flow codes\cite{mao_deepflame_2022,boivin2021benchmarking}, further study is required to reveal the fundamental physics involved in the benchmark. This would also help the code developers better target the implementation of the specific numerical algorithms. 

In the previous studies on TGV interacting with a flame (e.g. Refs\cite{abdelsamie_taylorgreen_2021,yang_evolution_2011}), only low Reynolds number conditions were considered, where laminar-to-turbulent flow transition was absent. In the present work, we perform several direct numerical simulation cases to investigate the interactions between hydrogen diffusion flames and the TGV at a range of Reynolds numbers. Instead of building a benchmark, this paper aims to enhance our understanding on the various dynamics of a hydrogen diffusion flame embedded in the Taylor-Green Vortex. 
The evolution of flame and vortex will be illustrated and the flame interacting turbulence at high Reynolds number will provide broader physical insight for this configuration for future numerical studies. Then the effects of molecular diffusion modelling (including mixture-averaged diffusion and unity Lewis number) on the flame structure\cite{cuenot_asymptotic_nodate} will be investigated as the differential diffusion plays a significant role role in a hydrogen-air combustion due to the high diffusivity of hydrogen and its atom radical\cite{han_thermal_2021}. In addition, all the DNS databases will be made publicly available, motivating the community building around database sharing\cite{domingo_recent_nodate,poinsot_applications_nodate}, which is essential for emerging research activities such as deep learning-assisted combustion modelling\cite{ZHANG2022112319,mao_deepflame_2022}.

The remainder of this paper is organised as follows. The methodology to conduct the DNS and configuration of the simulations are described in section 2 along with the in-house finite difference code used. Results are presented in section 3 and the effects of Reynolds number and non-unity Lewis number are analysed. Conclusions are summarized in the final section. 

\section{Numerical Method}

\subsection{DNS detail}
The governing equations solved for the conservation of mass, momentum, energy, and species mass fractions are written as
\begin{equation}
   \frac{\partial \rho}{\partial t}+\frac{\partial \rho u_i}{\partial x_i}  =0 
\end{equation}
\begin{equation}
   \frac{\partial \rho u_i}{\partial t}+\frac{\partial\left(\rho u_i u_j+p \delta_{i j}\right)}{\partial x_j} =\frac{\partial \sigma_{i j}}{\partial x_j} 
\end{equation}
\begin{equation}
    \frac{\partial \rho e}{\partial t}+\frac{\partial(\rho e+p) u_j}{\partial x_j} =\frac{\partial\left(\sigma_{i j} u_i-q_j\right)}{\partial x_j}+\dot{\mathcal{Q}}
\end{equation}
\begin{equation}
    \frac{\partial \rho Y_i}{\partial t}+\frac{\partial\left(\rho u_j Y_i\right)}{\partial x_j}=\frac{\partial\left(\rho D_i \frac{\partial Y_i}{\partial x_j} \right)}{\partial x_j}+\dot{\omega}_i
\end{equation}
where $u$ is the convective velocity, $\rho$ is the density, $p$ is the pressure, $Y_i$ is the $i$th species , $D_i$ is the coefficient of mass diffusion for species $i$, $\dot{\omega}_i$ is the mass based chemical source term of species $i$ and $\dot{\mathcal{Q}}$ is the external heat source term. The stress tensor and heat flux vector are expressed as
\begin{equation}
    \sigma_{i j}=\mu\left(\frac{\partial u_i}{\partial x_j}+\frac{\partial u_j}{\partial u_i}-\frac{2}{3} \delta_{i j} \frac{\partial u_k}{\partial x_k}\right)
\end{equation}
\begin{equation}
    q_i=-\lambda \frac{\partial T}{\partial x_i}+\rho \sum_{k=1}^N h_k Y_k V_{k, i}
\end{equation}
where $\mu$ is the fluid viscosity, $\lambda$ is the coefficient of thermal conductivity, $h_k$ is the enthalpy of species $k$.

The above governing equations are solved using an in-house finite-difference code, Advanced flow Simulator for Turbulence Research (ASTR), which has been used for several previous studies\cite{fang_improved_2019,fang_optimized_2013,fang_turbulence_2020}. The convective terms are solved in the skew-symmetric form and the diffusive terms are solved in the Laplacian form. A sixth-order explicit central scheme (ECS6)\cite{fang_improved_2019} is applied on the calculation of the first and second derivatives, which are discretized on a uniform spatial grid. The chemical source terms are calculated using the chemistry ODE solver through the Cantera Fortran interface. The time integration is conducted using a three-step total variation diminishing third-order Runge-Kutta scheme. The 9-species mechanism of Boivin et al. \cite{BOIVIN2011517} is used for the hydrogen combustion kinetics. As for the molecular diffusion model, the mixture-averaged diffusion is applied in Case 1 to 4 (see Table~\ref{table:1}) and the unity Lewis number (UL) is applied in Case 1-UL to 4-UL. For the UL cases, the diffusion coefficient $D_i$ is identical for all species such that $D_i = \frac{\lambda}{\rho C_p}$, where $C_p$ is the specific heat capacity at constant pressure. For the cases with mixture-averaged diffusion, the diffusion coefficient is computed as 
\begin{equation}
    D_i = \frac{1-Y_i}{\sum_{j\neq i} X_j/D_{ji}}
\end{equation}
where $D_{ji}$ are the binary diffusion coefficients.

\subsection{TGV-flame configuration}
Following the work of Abdelsamie et al.\cite{abdelsamie_taylorgreen_2021}, the computational domain is a cubic box of size $[0,L]^3$ with $L=2\pi L_0$. The central part of the box is initially filled with a H$_2$/N$_2$ mixture (fuel region where $Y_{\mathrm{H}_2}^0 = 0.0556$, $T = 300$~K) while the remaining part is filled with air with $Y_{\mathrm{O}_2}^0 = 0.233$. The shape steps at the interface separating different regions are smoothed out using hyperbolic tangent functions:
\begin{equation}
    R_d(x)=|x-0.5 L|
\end{equation}
\begin{equation}
    \psi(x)=0.5\left[1+\tanh \left(\frac{c\left(R_d(x)-R\right)}{R}\right)\right]
\end{equation}
where $R = L/8 $ and $c = 3 $ are the half-width of the central slab and stiffness parameter, respectively. The species mass fraction distribution can be expressed as $Y_{\mathrm{H}_2}(x)=Y_{\mathrm{H}_2}^0(1-\psi(x))$, $Y_{\mathrm{O}_2}(x)=Y_{\mathrm{O}_2}^0 \psi(x)$, $Y_{\mathrm{N}_2}(x)=1-Y_{\mathrm{H}_2}(x)-Y_{\mathrm{O}_2}(x)$. The equilibrium temperature at each point is calculated for the local mixture for constant pressure and enthalpy using \textit{libcantera} functions along with the thermodynamic and transport properties specified for the initial field.

The initial condition of the 3-D TGV is given by 
\begin{equation}
    u(x,y,z) = u_0\sin (\frac{x}{L_0}) \cos (\frac{y}{L_0}) \cos (\frac{z}{L_0})
\end{equation}
\begin{equation}
    v(x,y,z) = -u_0\cos (\frac{x}{L_0}) \sin(\frac{y}{L_0}) \cos (\frac{z}{L_0})
\end{equation}
\begin{equation}
    w(x,y,z) = 0
\end{equation}
\begin{equation}
    p(x,y,z) = p_0 +\frac{\rho_0 {u_0}^2}{16}[\cos (\frac{2x}{L_0})+\cos (\frac{2y}{L_0})][\cos (\frac{2z}{L_0})+2]
\end{equation}
where $p_0$ is the atmospheric pressure and $\rho_0$ is the standard density calculated using the ideal-gas law with $p_0$ and $T_0 = 300$~K. Differing to Ref.\cite{abdelsamie_taylorgreen_2021}  which mainly focused on a laminar reacting flow ($Re \approx 250$, see validation in the Appendix), we consider situations with higher turbulence intensity through increasing the initial velocity magnitude $u_0$ and vortex length scale $L_0$. Table~\ref{table:1} provides the numerical configuration and relevant parameters for the DNS cases considered in this work. The different velocity magnitude $u_0$ and integral scale $L_0$ lead to different evolution of the flame structure and vortex field. A flow reference time is defined as $\tau_{f}=L_0/u_0$, indicating the evolution of the TGV. The initial Reynolds number of the configuration is defined using the minimal viscosity $\nu_{min} = 1.5897\times 10^{-5}$ m$^2$/s obtained for air at $300$~K. In order to assist the investigation of the vortex-flame interaction, a chemical reference time is defined using the ignition delay time of a H$_2$/O$_2$ mixture ($Y_{H2}=0.0556,Y_{O2}=0.233$) at 1200~K, $\tau_{chem} = 0.05$~ms, which indicates the evolving state of the flame in the cases. It is noteworthy that the flow reference time defers in different cases while the chemical reference time remains constant. In the following sections, the evolution of the flame structure and vortices will be analysed and compared based on these two time scales. 

\begin{table}[htbp]
\begin{tabular}{lllll}
\hline
Case                 & 1/1-UL      & 2/2-UL      & 3/3-UL     & 4/4-UL  \\ \hline
Velocity magnitude $u_0$ (m/s)          & 12.5            & 25.0            & 50.0        & 25.0    \\
Integral scale $L_0$ (mm)             &   1.0         &   1.0         &    1.0  &       2.0  \\
Initial Reynolds number Re  &  800           &    1600         &  3200    & 3200       \\
Kolmogorov scale $\eta_k$ ($\mu$m)         &  41.7           &       24.8       &   14.8   &   29.8               \\
Flow ref. time $\tau_{f}$ (ms)    & 0.08            & 0.04             & 0.02        &0.08    \\
Ratio $\tau_{f}/\tau_{chem}$  & 1.6            & 0.8            & 0.4        &1.6    \\
Discretized grids $N^3$          & $256^3$ & $512^3$ & $512^3$ & $512^3$\\
Grid resolution $\Delta x$ ($\mu$m)       &  24.5          &       12.3      &   12.3      &  24.5      \\
\hline
\end{tabular}
\caption{\label{table:1} Numerical configuration for the DNS cases.}
\end{table}

\begin{figure}[htbp]
\centering
\includegraphics[width=0.9\linewidth]{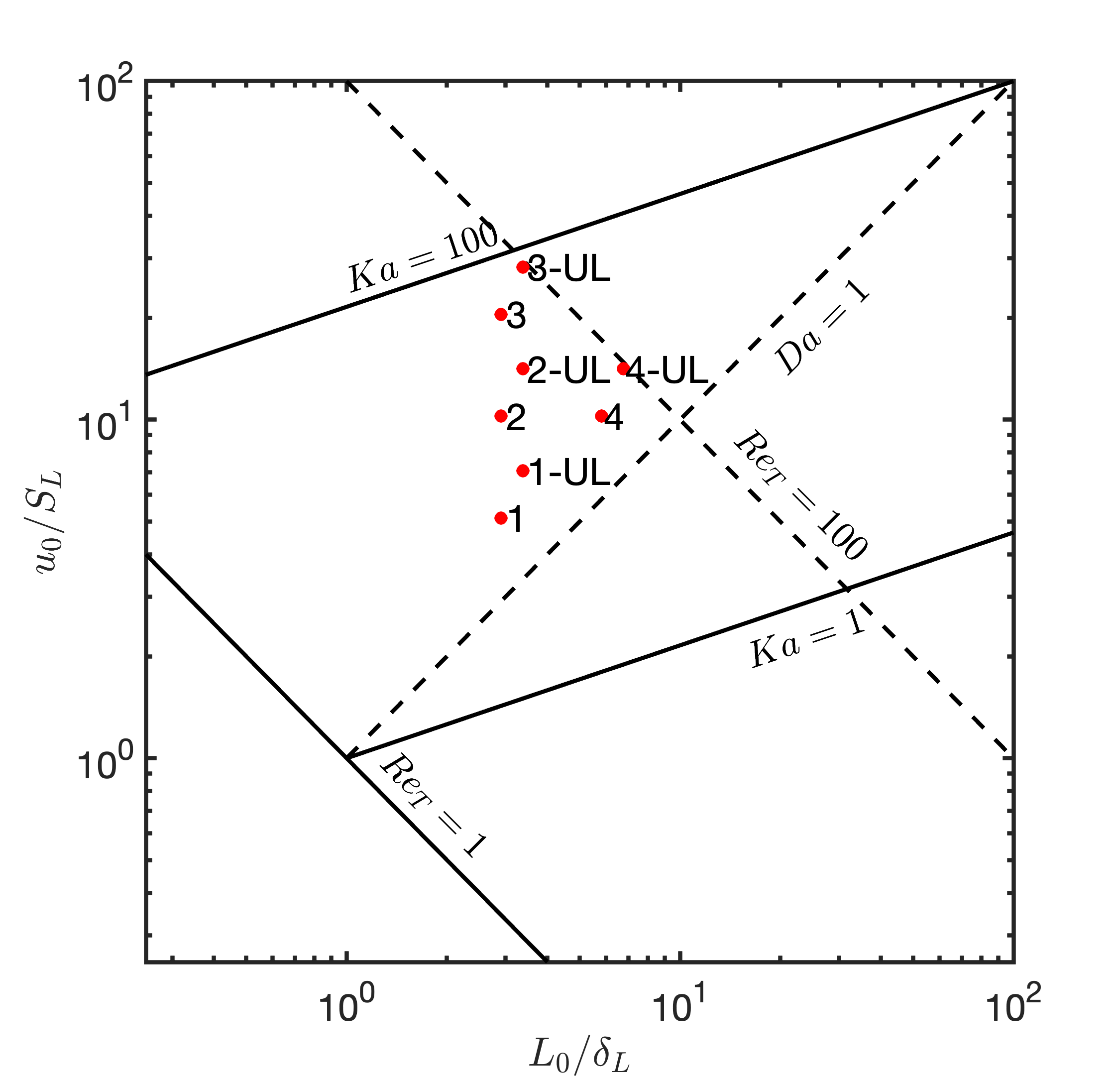}
\caption{Diagram of regimes showing the location of the DNS cases. Case 1 to Case 4 refer to cases with mixture-averaged model and Case 1-UL to Case 4-UL with unity Lewis number.}
\label{fig1}
\end{figure}

Two sets of four DNS runs were performed with mixture-averaged and unity-Lewis models, totalling 8 simulation cases. For Case 1/1-UL to Case 3/3-UL, $u_0$ doubles progressively with constant $L_0$, resulting in $Re={u_0L_0}/\nu_{min}=800,1600,3200$. In Case 4/4-UL, $u_0$ and $L_0$ double at the same time compared to Case 1/1-UL leading to an equal $\tau_f$, but with a 4 times Reynolds number $Re = 3200$. This configuration with a larger computing domain aims at exploring a case with higher turbulence intensity while keeping $\tau_f$ constant. Furthermore, the influence of the flame-induced effects can be studied through comparing Case 3/3-UL with Case 4/4-UL as they have the same Reynolds number. Figure~\ref{fig1} presents the classic regime diagram for turbulent combustion and the location of the DNS cases are marked. Here the laminar flame thickness is calculated using stoichiometric condition at 300 K resulting in $\delta_L = 0.344$~mm for the mixture-averaged cases, and $\delta_L = 0.296$~mm for the unity Lewis cases. The stoichiometric laminar flame velocity is calculated in a similar manner and it is $S_L = 2.44$~m/s for the mixture-averaged cases and $S_L = 1.77$~m/s for the unity Lewis cases. Thus, the turbulent Reynolds number depicted in the figure is defined as $Re_T = (u_0/S_L)(L_0/\delta_L)$. The Damkohler number is defined as $Da = (L_0/u_0)/(\delta_L/S_L)$ and the Karlovitz number is defined as $Ka = (\delta_L/L_0)^{1/2}(u_0/S_L)^{3/2}$.

The time step of the simulation is $\Delta t = 1.25\times10^{-8}$~s for Case 1/1-UL whose computing domain is discretized with a $256^3$ uniform mesh, while it is $\Delta t = 6.25\times10^{-9}$~s  for Case 2/2-UL to 4/4-UL with a $512^3$ uniform mesh. A mesh sensitivity study was conducted showing that for $Re=800$ and larger, this benchmark requires a finer mesh to resolve the small-scale turbulence-flame interactions. The time steps chosen are small enough to capture the full dynamics of fast reacting radicals such as H, O and OH. In Case 4/4-UL, the flame is resolved by $\delta_L/\Delta x > 20$ grid points, resulting in a good resolution for the flame structure. All turbulence structures are adequately resolved using a grid resolution $\Delta x$ smaller than the Kolmogorov scale $\eta_k$ in the cases as listed in Table 1. Considering the computing cost of the simulations using ASTR, the
Reduced Computational Time (RCT) is defined with the expression: RCT = TCPU/($N$it × $N$p). The variable TCPU represents the total CPU time for the simulation and it is computed by multiplying $N$cores and the total CPU time. Here $N$p is the number of grid points, $N$cores is the CPU core number, $N$it is the iteration steps. RCT is introduced to evaluate the CPU time needed to simulate one time-step on a single degree of freedom by using one processor. Its value is calculated as RCT = 60 $\mu$s for the simulations. The CPU used here is AMD EPYCTM 7742 64-core 2.25 GHz processor. 

\begin{figure}[htbp]
\centering
\includegraphics[width=0.9\linewidth]{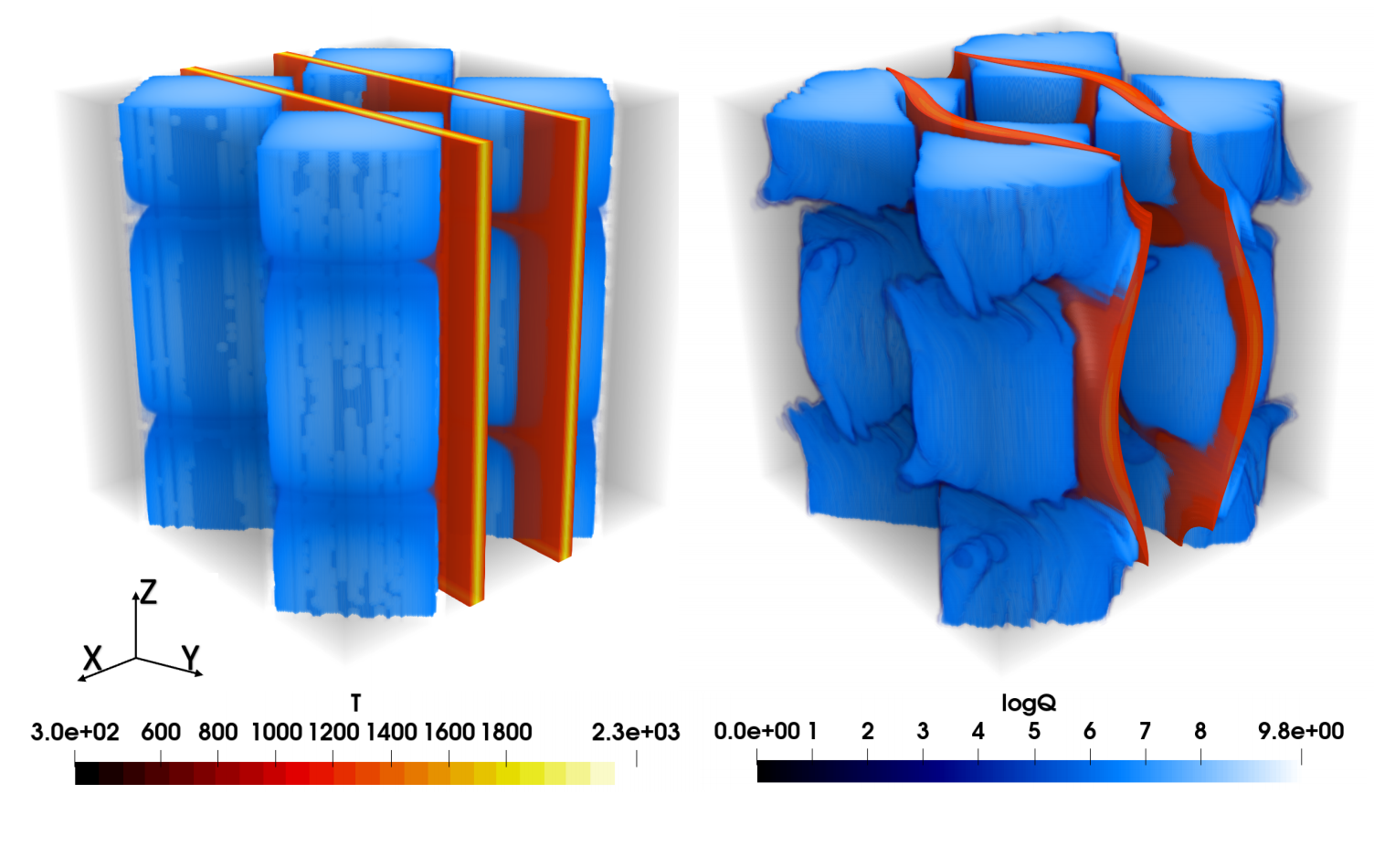}
\caption{Left: Initial field of temperature and log scale of Q-criterion. Right: Field at $t=0.5\tau_f$ for Case 1, where $t=0.04$~s.}
\label{fig2}
\end{figure}

Figure~\ref{fig2}a shows the initial field of Case 1, where the temperature field is displayed in surface rendering and the vortex is visualized through the log scale of Q-criterion displayed in volume rendering. The steep profile of temperature distribution simulates that of a real flame. All cases listed in Table~\ref{table:1} are initialised similarly with differences only in the TGV velocity and length scale. Figure~\ref{fig2}b shows the evolved field at $0.5\tau_f$ after the initial state, where the flame surface starts to bend under the influence of vortex motion. Subsequent evolution and comparison among different cases will be investigated and evaluated further in the next section.

\section{Results and Discussion}
\subsection{Evolution of the flame and vortex}
The evolution of the flow field is driven by both flame-induced and vortex-induced motions.  As the TGV with cold flow has already been widely studied in nondimensional forms~\cite{fang_improved_2019,yang_evolution_2011}, here we mainly focus on the TGV-flame interactions and the comparisons among the different cases, especially the turbulent state in cases with higher Reynolds number. In Subsection A, the evolution of the flame and vortex in Case 1 to Case 3 are first compared and discussed, while the effects of unity Lewis number and vortex length scale will be investigated in Subsection B and C.

\begin{figure}[htbp]
\centering
\includegraphics[width=\linewidth]{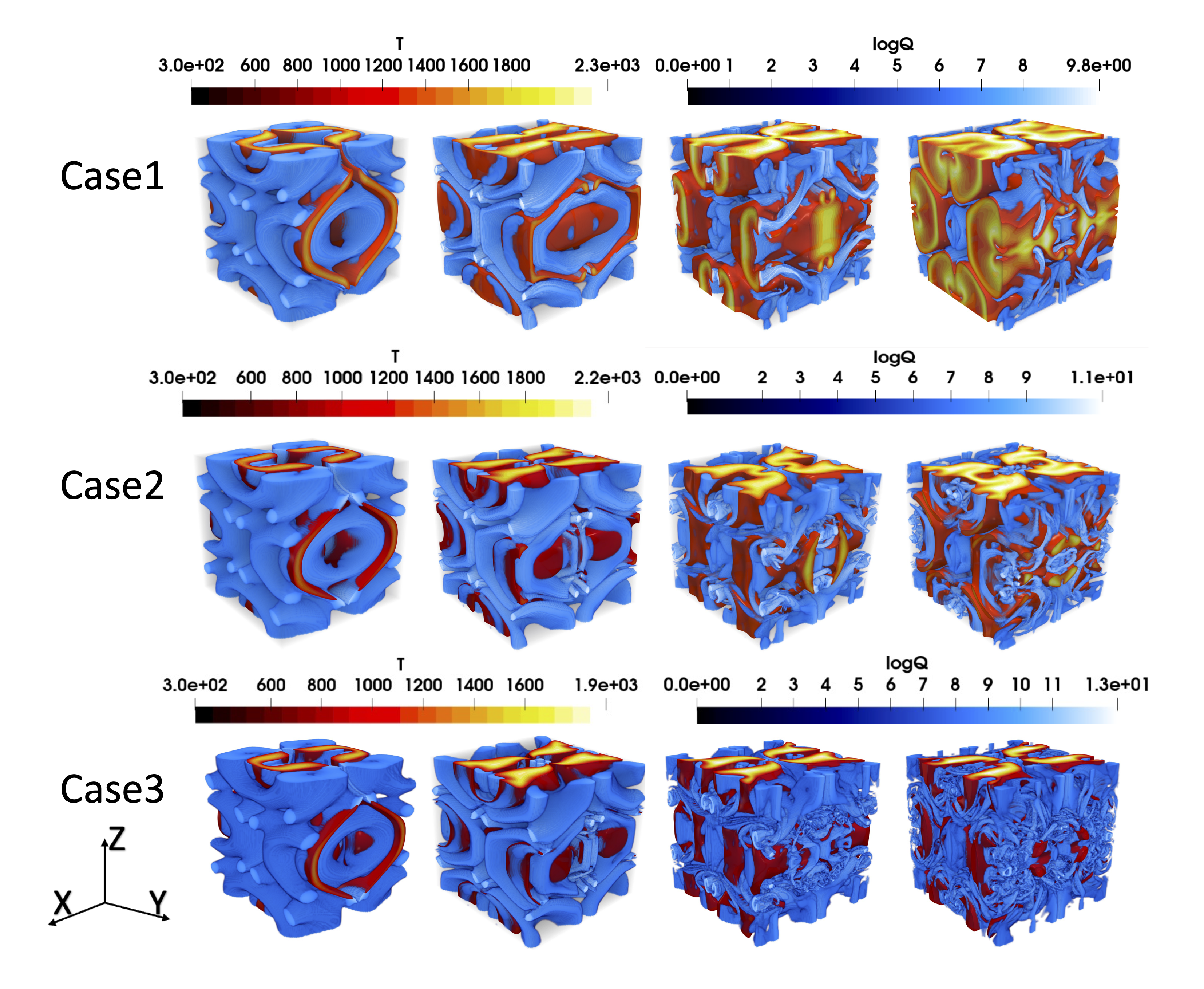}
\caption{Evolution of the temperature field and log-scale of the Q-criterion field for the 3 cases. In each line, the fields are displayed at $t=2\tau_f, 4\tau_f, 6\tau_f, 8\tau_f$ for each case, where the colormaps of temperature and log-Q are given accordingly.}
\label{fig3}
\end{figure}

In Fig.~\ref{fig3}, the temperature field and log scale of Q-criterion field are shown at $t=2\tau_{f}, 4\tau_{f}, 6\tau_{f}, 8\tau_{f}$ for each case, in order to ensure that the vortex rotating in different cases reach the same state at the same relative time. It is worth reminding that $\tau_f$ differs in the 3 cases (see Table~\ref{table:1}). It can be seen that the vortices have very similar dynamics typically at $t=2\tau_{f}, 4\tau_{f}$ for the three cases, where the initial lumpy vortices are rolled up into vortex tubes surrounded by the spreading flame surface. The behaviors of the vortices start to differ in the following time steps. In Case 1, parts of the vortex tubes are stretched into thinner tubes at $t=6\tau_{f}$, while they start to dissipate under the effects of strong heat release at $t=8\tau_{f}$. In Case 2, due to the higher $Re$ the vortex tubes are stretched and twisted into even thinner ones at $t=6\tau_{f}$. It can also be observed that fine vortex filament structures start to appear in the middle of the domain where the strongest stretching occurs. At $t=8\tau_{f}$, some helical geometry and turbulence-like flow structures can be observed. Compared to non-reacting TGV with Re=1600\cite{fang_improved_2019}, the TGV-flame Case 2 shows relatively weaker turbulence because of the suppression effect caused by heat release of chemical reactions. In Case 3, the vortex tubes are constantly stretched, split and twisted under strong turbulence intensity from $t=4\tau_{f}$ to $t=8\tau_{f}$. It is demonstrated from the figure that small-scale structures are generated all over the domain at $t=8\tau_{f}$, which indicates a fully evolved turbulent state. In addition to the $Re$ number effects, it should also be noted that from Case 1 to 3, the flame evolution is lagged due to smaller $\tau_{f}$ leading to a relatively weaker heat release at the same $t/\tau_{f}$ instant. 

\begin{figure}[htbp]
\centering
\includegraphics[width=0.7\linewidth]{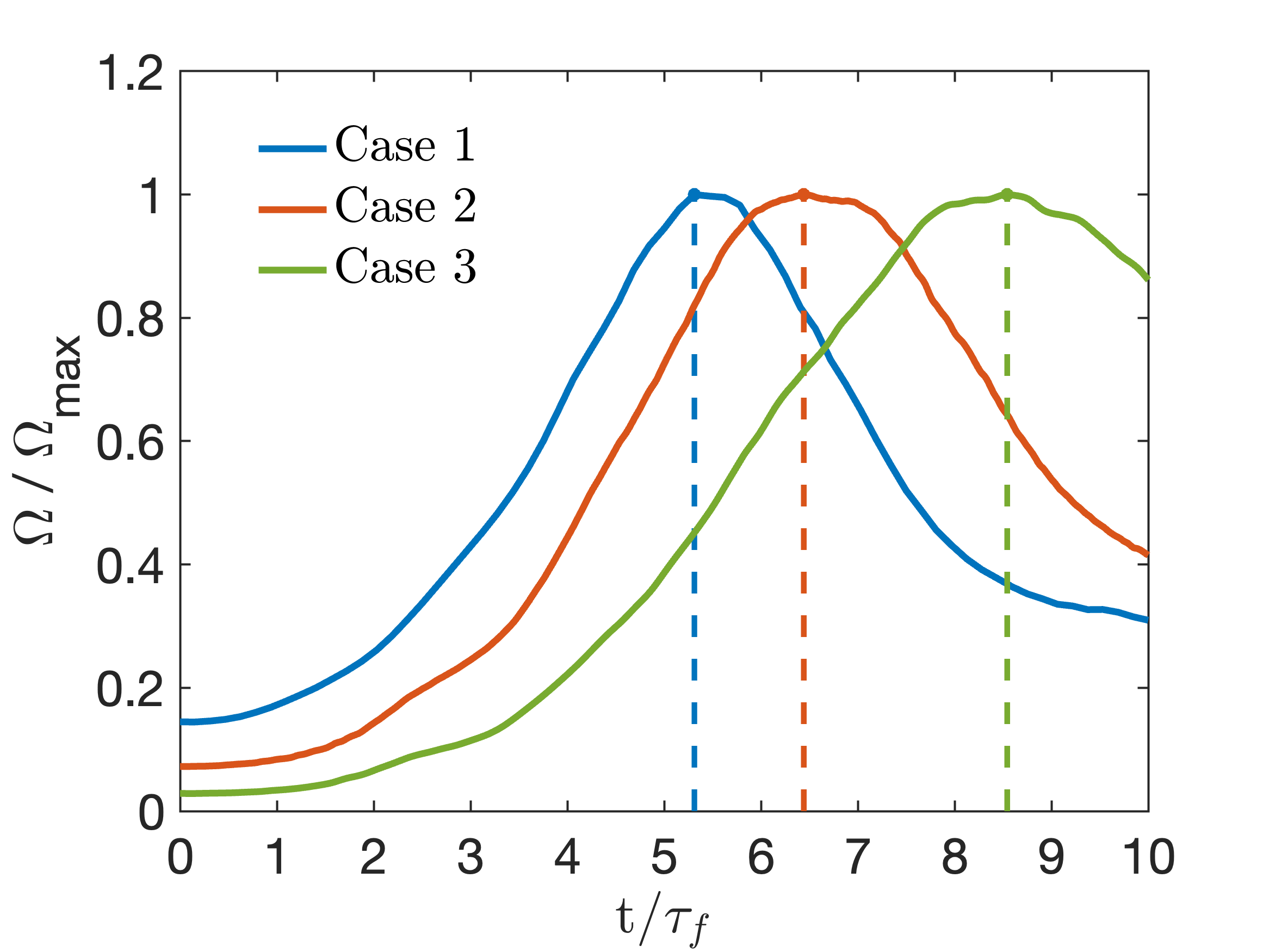}
\caption{Temporal evolution of the volume-averaged enstrophy scaled by $\tau_f$ for each case. The relative time at which enstrophy reaches its peak value is marked using dashed lines.}
\label{fig4}
\end{figure}

\begin{figure}[htbp]
\centering
\includegraphics[width=0.7\linewidth]{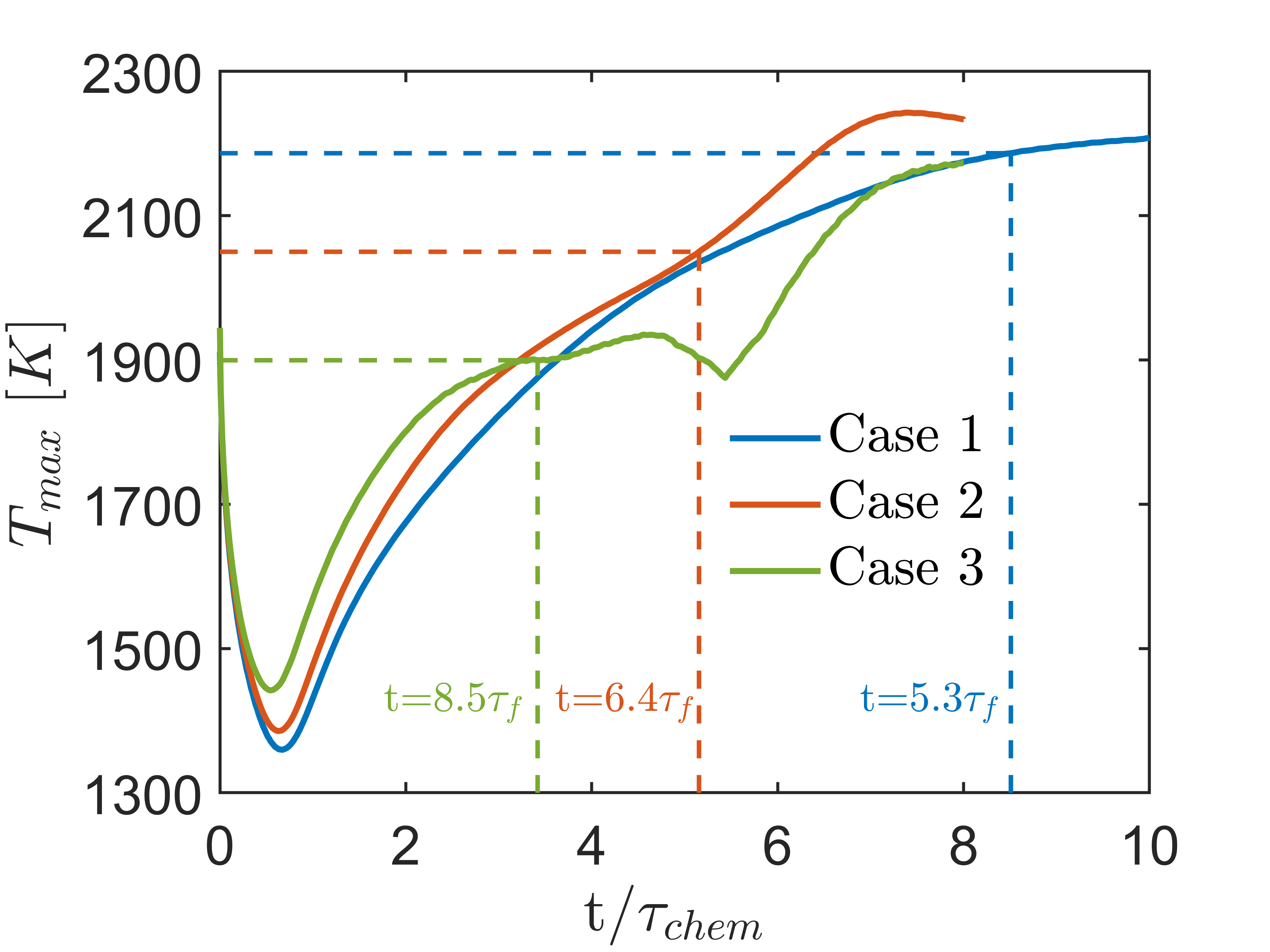}
\caption{Temporal evolution of the maximum temperature for the 3 cases scaled by $\tau_{chem}$. The time of enstrophy reaching its peak value scaled by $\tau_f$ is marked in the figure scaled by $\tau_{chem}$ for each case using dashed lines.}
\label{fig5}
\end{figure}

Figure \ref{fig4} shows the evolution of the volume-averaged enstrophy in different cases using the flow reference time $\tau_{f}$. As seen in the figure, the relative time for enstrophy reaching the peak value is delayed with the increasing $Re$, i.e. larger velocity magnitude. In Case 1, the enstrophy of the system reaches the peak value at $5.3\tau_{f}$, while it is about $6.4\tau_{f}$ and $8.5\tau_{f}$ in Case 2 and Case 3, respectively. Figure~\ref{fig5} depicts the temporal evolution of the maximum temperature for the 3 cases scaled by $\tau_{chem}$. Corresponding to Fig. \ref{fig4}, the time for enstrophy reaching its peak value scaled by $\tau_f$ is marked for each case. This is to compare the state of the flame at specific TGV flow evolution times. It is widely acknowledged that the heat release generated by chemical reaction has suppression effects on the turbulence intensity\cite{poinsot_theoretical_2005}. Combining Figs.~\ref{fig3} and \ref{fig5}, it can be observed that the strength of the flame is weaker and the maximum temperature is lower for cases with higher turbulence intensity at the same relative time $\tau_{f}$, which means the suppression effects of chemical reacting are relatively delayed resulting in a longer evolution time before reaching the peak value of enstrophy.

Regarding the hydrodynamic effects on the flame, it is illustrated in Fig. \ref{fig5} that the maximum temperature is higher for cases with larger velocity magnitude in the early stage of the evolution, which ranges from $t=0$ to $3\tau_{chem}$. This is because the significant intensity of the vortices provides abundant kinetic energy and accelerates the mixing of the mixture leading to an earlier ignition. Note that in the initial field no fuel is present in the high temperature region. From $t=3\tau_{chem}$ to $5\tau_{chem}$, it is shown that the bending of the temperature curve becomes more apparent for cases with high turbulence intensity because of the higher strain rate induced by the flow. It can also be observed in Fig. \ref{fig3} that the higher strain rate leads to larger curvatures of the flame surface for cases with higher velocity magnitude. In Fig. \ref{fig5}, the sudden change of the maximum temperature at around $t=5.7\tau_{chem}$ in Case 3 is probably due to the change of position for the fluid parcel where $T_{max}$ is located in the field. This speculation requires further investigation via a Lagrangian-type of tracing analysis, which will be conducted in a future work.

\begin{figure*}[htbp]
    \centering
    \subfigure[Case 1 at $t=5.3\tau_f$]{
		\begin{minipage}[t]{0.3\linewidth}
			\centering
			\includegraphics[width=\linewidth]{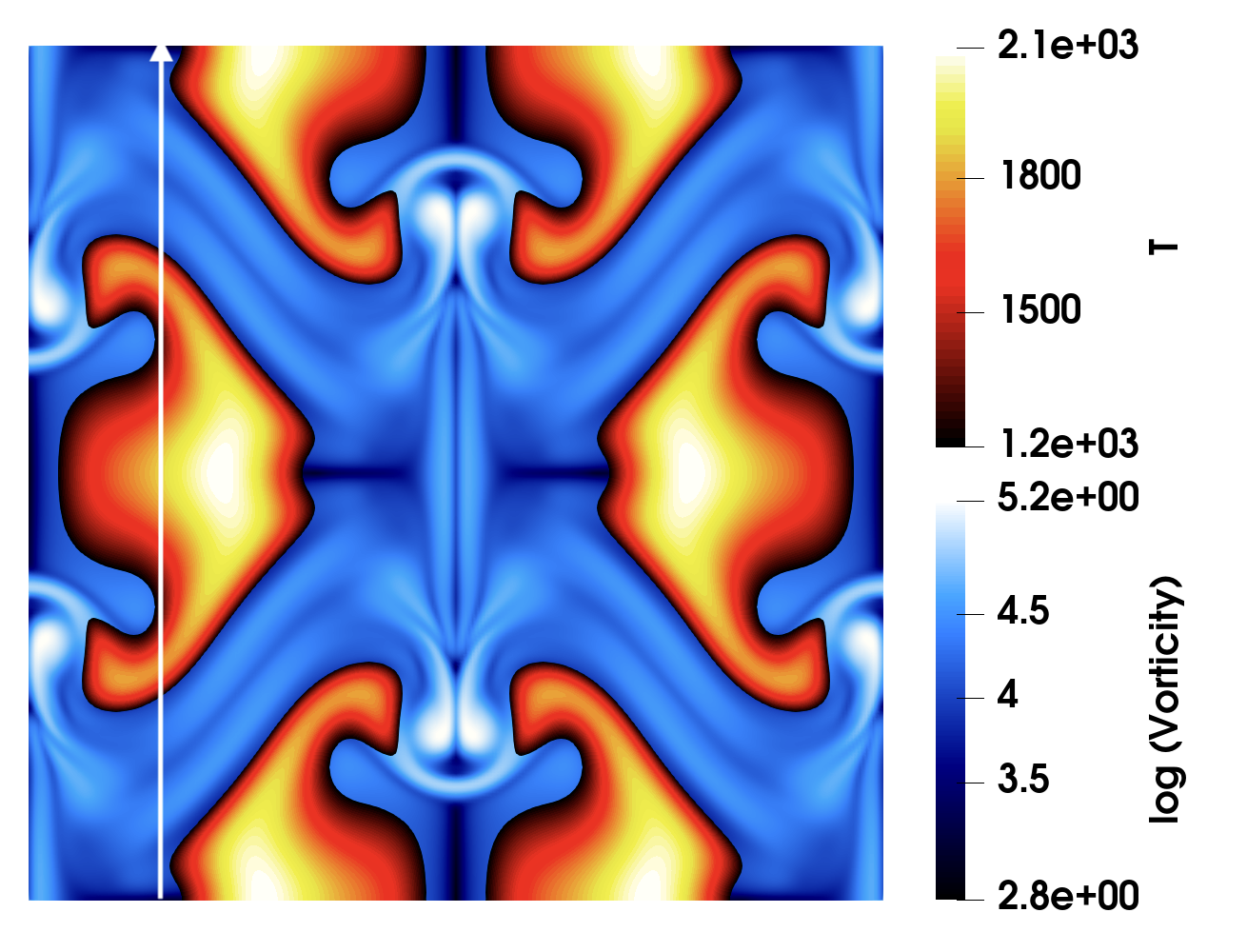}
		\end{minipage}
    }
    \subfigure[Case 2 at $t=6.4\tau_f$]{
		\begin{minipage}[t]{0.3\linewidth}
			\centering
			\includegraphics[width=\linewidth]{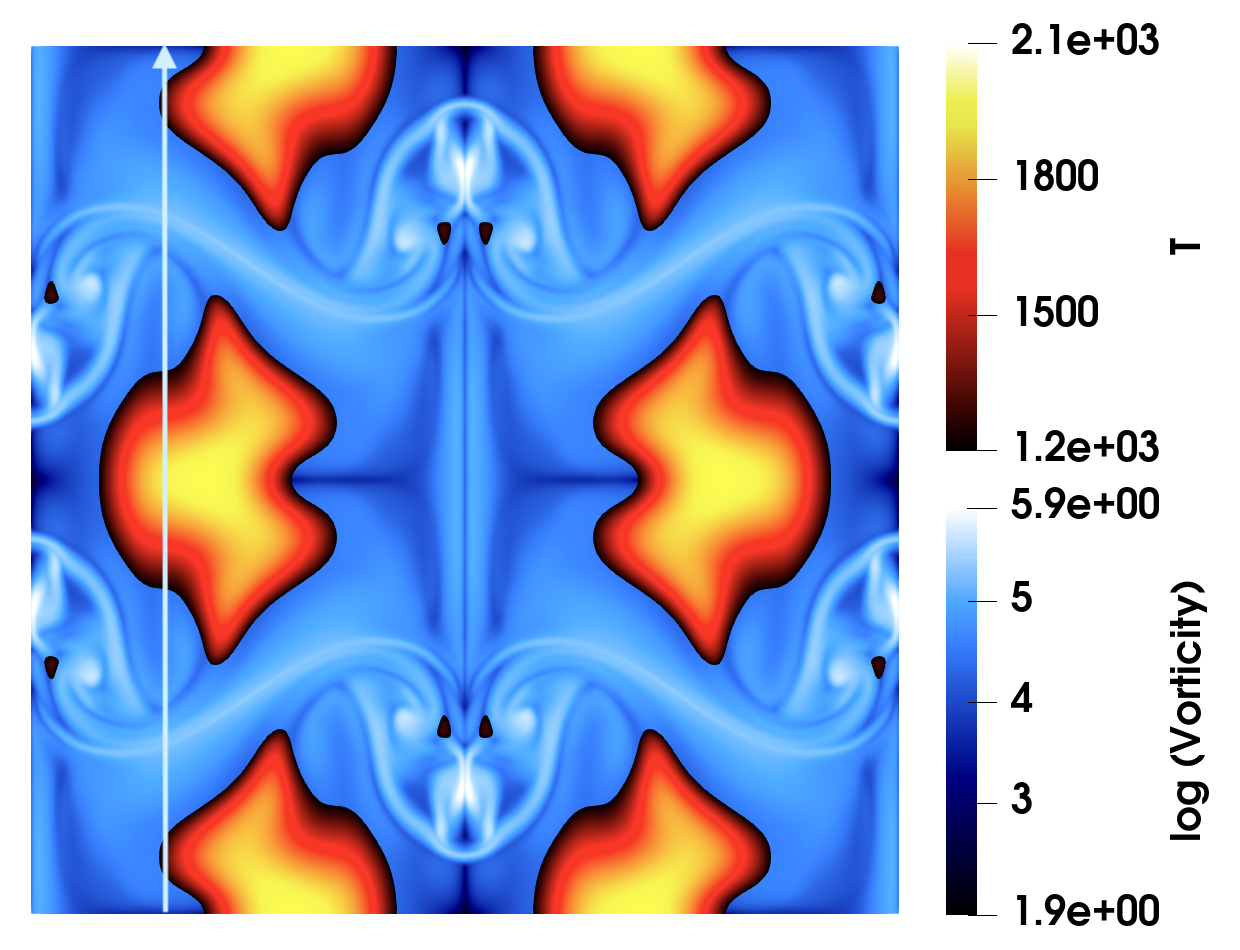}
		\end{minipage}
    }
    \subfigure[Case 3 at $t=8.5\tau_f$]{
		\begin{minipage}[t]{0.3\linewidth}
			\centering
			\includegraphics[width=\linewidth]{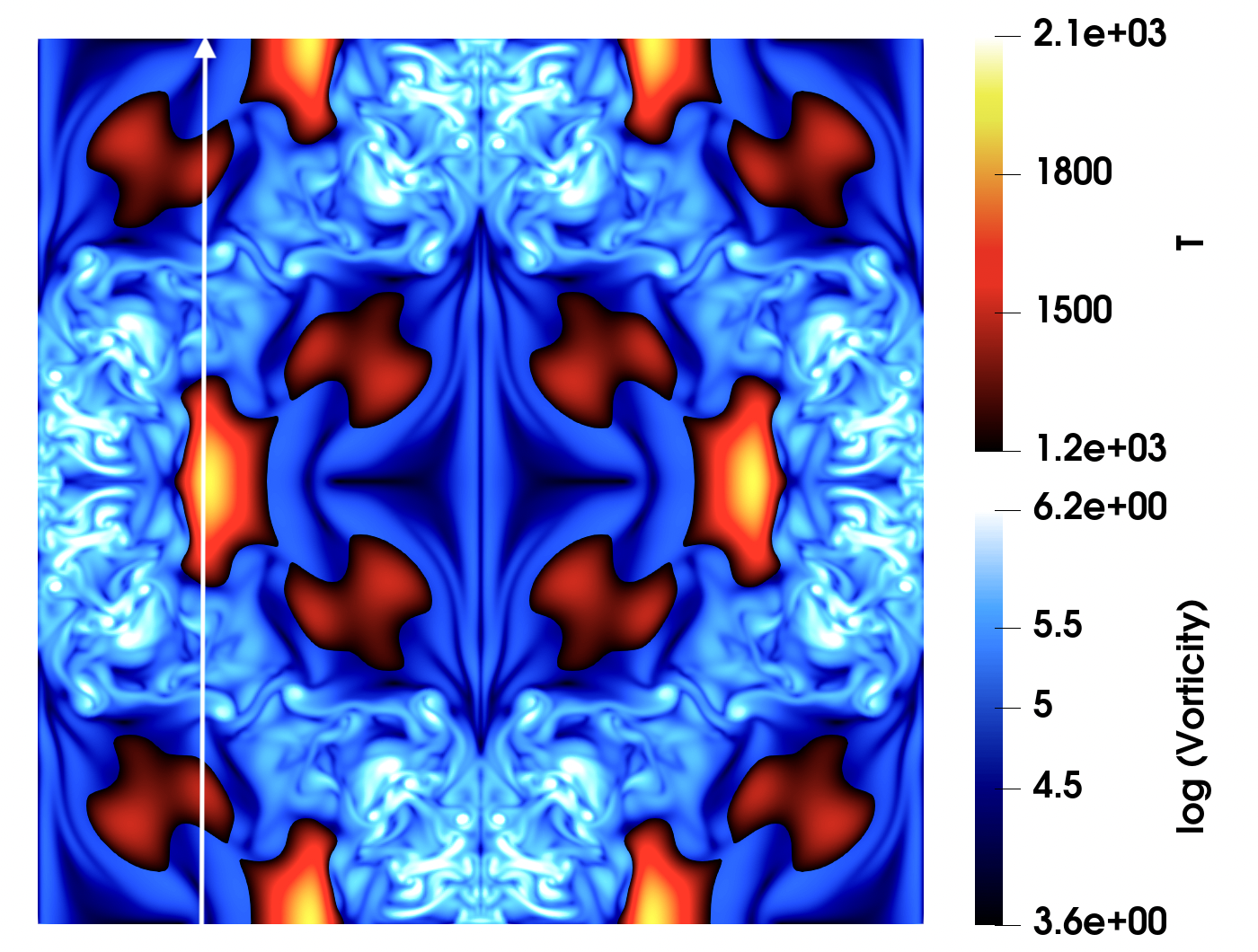}
		\end{minipage}
    }
    \centering
	\caption{Temperature and log (Vorticity) field in the $x = L/6, y-z$ plane at which time enstrophy reaching the peak value for the 3 cases. The white arrow line along z-direction in each subfigure is chosen crossing the combustion region in each case.}
	\label{fig6}
\end{figure*}

For a close analysis of the vortex-flame interaction and confirmation of the above discussion, 2D-slices in the $x = L/6,\,y-z$ plane are shown in Fig. \ref{fig6} for the time instant at which enstrophy reaches the peak value in each case. In this figure, the flame structure is visualised via the temperature field and the vortex structure is displayed in the log (vorticity) field. The slices are chosen at a specific plane rather than the mid-plane for the reason that this profile depicts the significant interaction between the flame and vortex, while there is little vortex motion on the mid-plane due to the characteristic structure of the TGV. It is illustrated in Fig. \ref{fig6}(a) that the the flame is quite strong at $t=5.3\tau_f$ (when enstrophy reaches its peak value) for Case 1, while the vortices are mainly presented as vortex tubes in large scale under the effects of strong heat release. Compared to Case 1, the combustion region is limited and the overall temperature is lower in the region for Case 2 and Case 3. Figure~\ref{fig6}(b) shows that helical structures of the vortices appear at $t=6.4\tau_f$ for Case 2 and the flame is absent in these areas. In Fig.~\ref{fig6}(c) for Case 3, a fully evolved turbulent state is shown and the breaking and quenching of the flame can be observed due to the intense stretching effects. 

\begin{figure*}[htbp]
    \centering
    \subfigure[Case 1]{
		\begin{minipage}[t]{0.3\linewidth}
			\centering
			\includegraphics[width=2.1in]{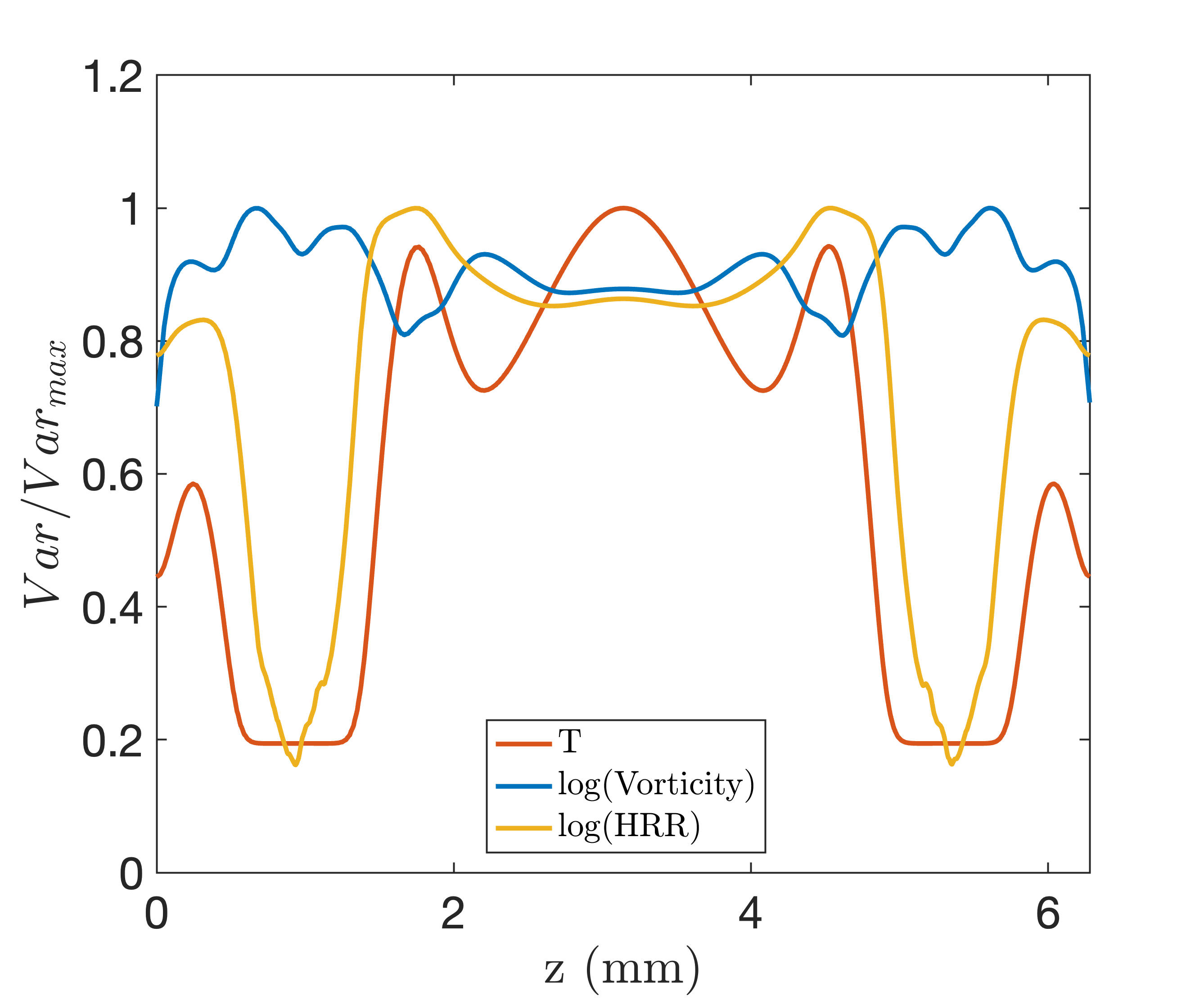}
		\end{minipage}
    }
    \subfigure[Case 2]{
		\begin{minipage}[t]{0.3\linewidth}
			\centering
			\includegraphics[width=2.1in]{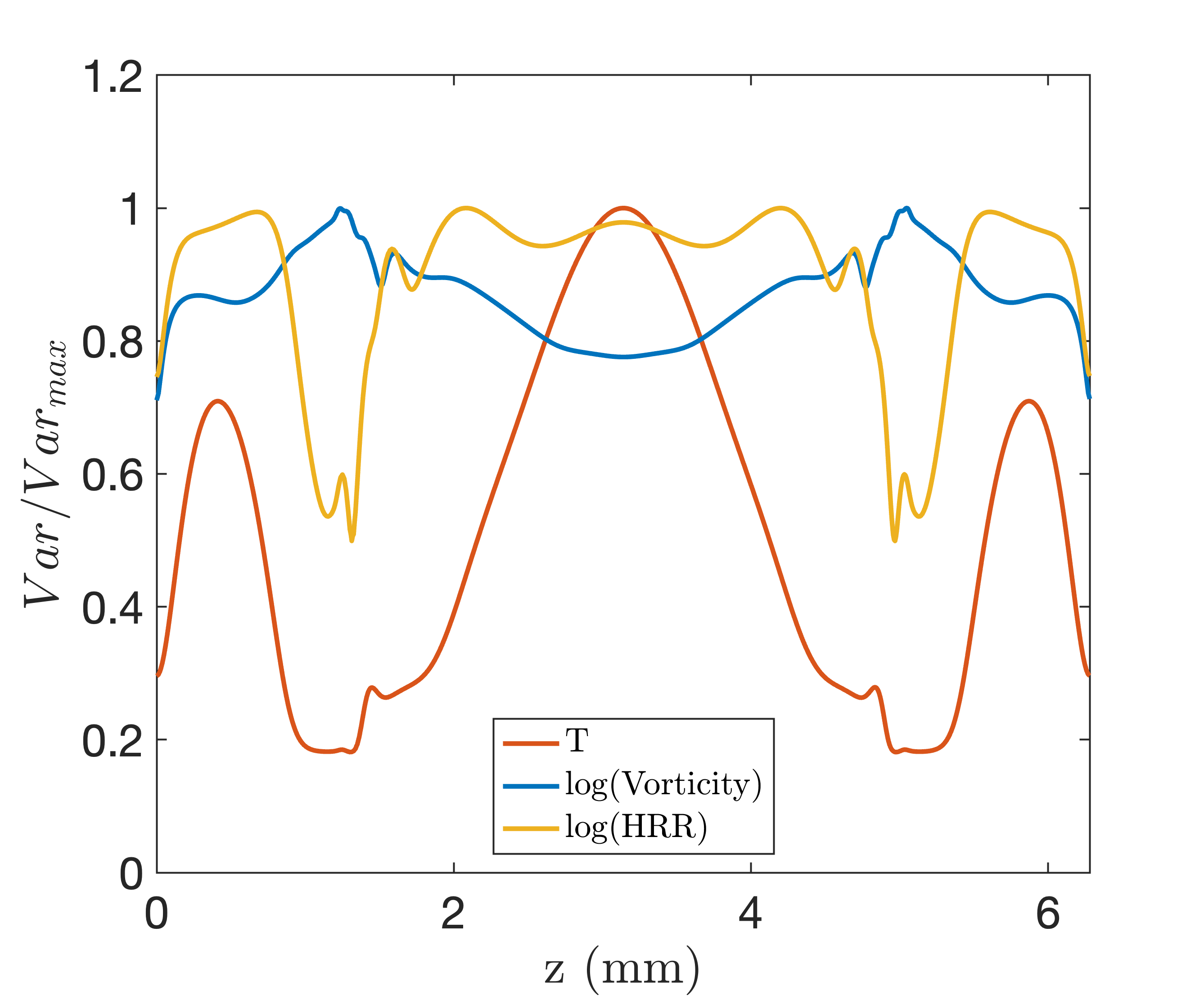}
		\end{minipage}
    }
    \subfigure[Case 3]{
		\begin{minipage}[t]{0.3\linewidth}
			\centering
			\includegraphics[width=2.1in]{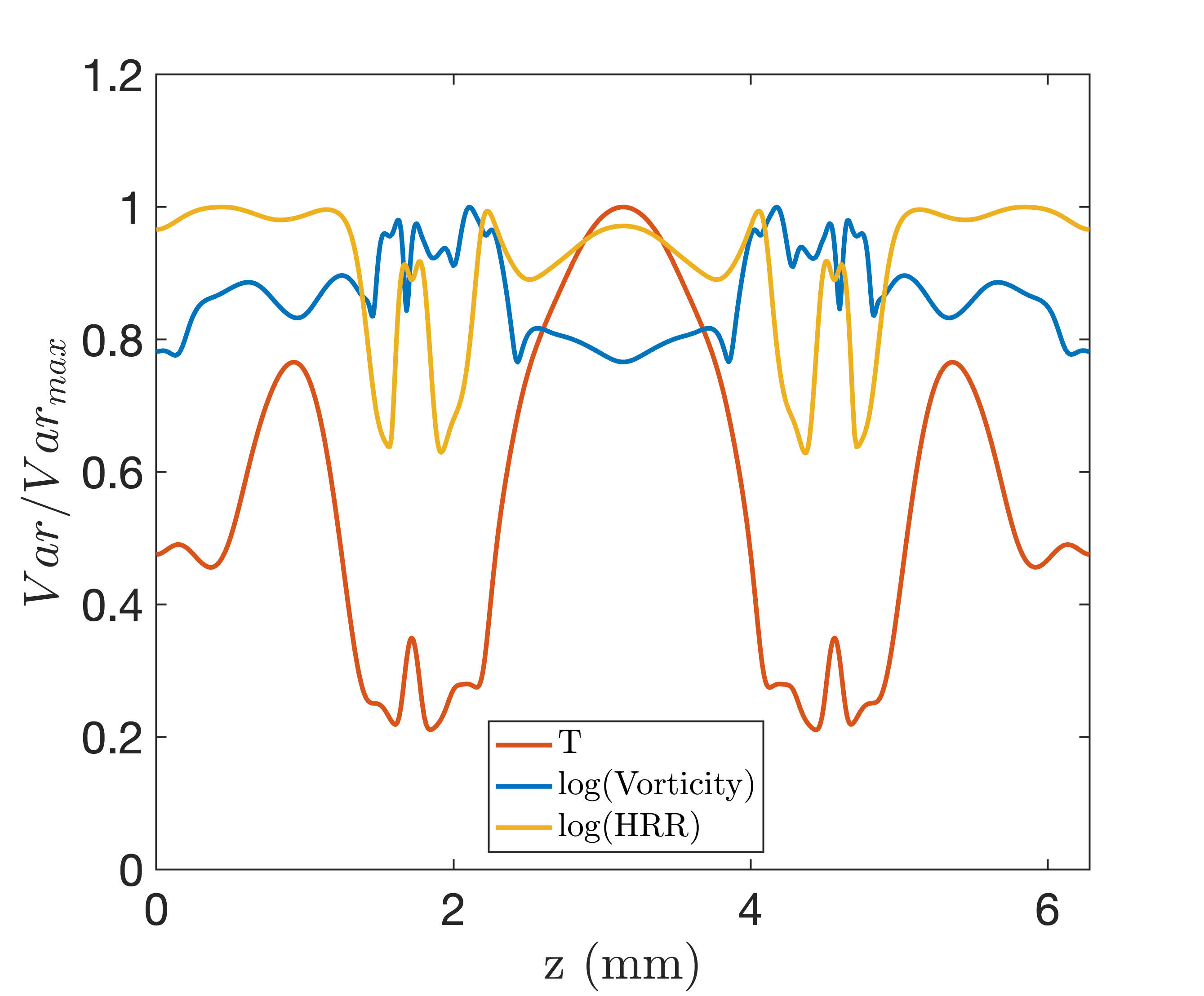}
		\end{minipage}
    }
    \centering
	\caption{Normalised temperature, vorticity and heat release rate along the white arrow lines marked in Fig. \ref{fig6}.}
	\label{fig7}
\end{figure*}

In order to further investigate the relationship between the vorticty field and heat release, specific white arrow lines are marked in Fig. \ref{fig6} crossing the combustion region along y-direction for each case. Normalized temperature, log (Vorticity), log (Heat Release Rate) along each line are calculated and depicted in Fig. \ref{fig7}. Despite the differences of the flame structure among the cases, it is similar in the 3 subfigures that vorticity decreases with the increasing of temperature and heat release rate, and \textit{vice versa}. The vorticity value falls to a local minima at which position the temperature and heat release rate maintain a high level. This common behavior depicted in Fig. \ref{fig7} confirms that the heat release has suppression effects on the vortex intensity. 

\subsection{Effects of diffusion model on the flame}
\begin{figure*}[htbp]
    \centering
    \subfigure[Case 1]{
		\begin{minipage}[t]{0.45\linewidth}
			\centering
			\includegraphics[width=3in]{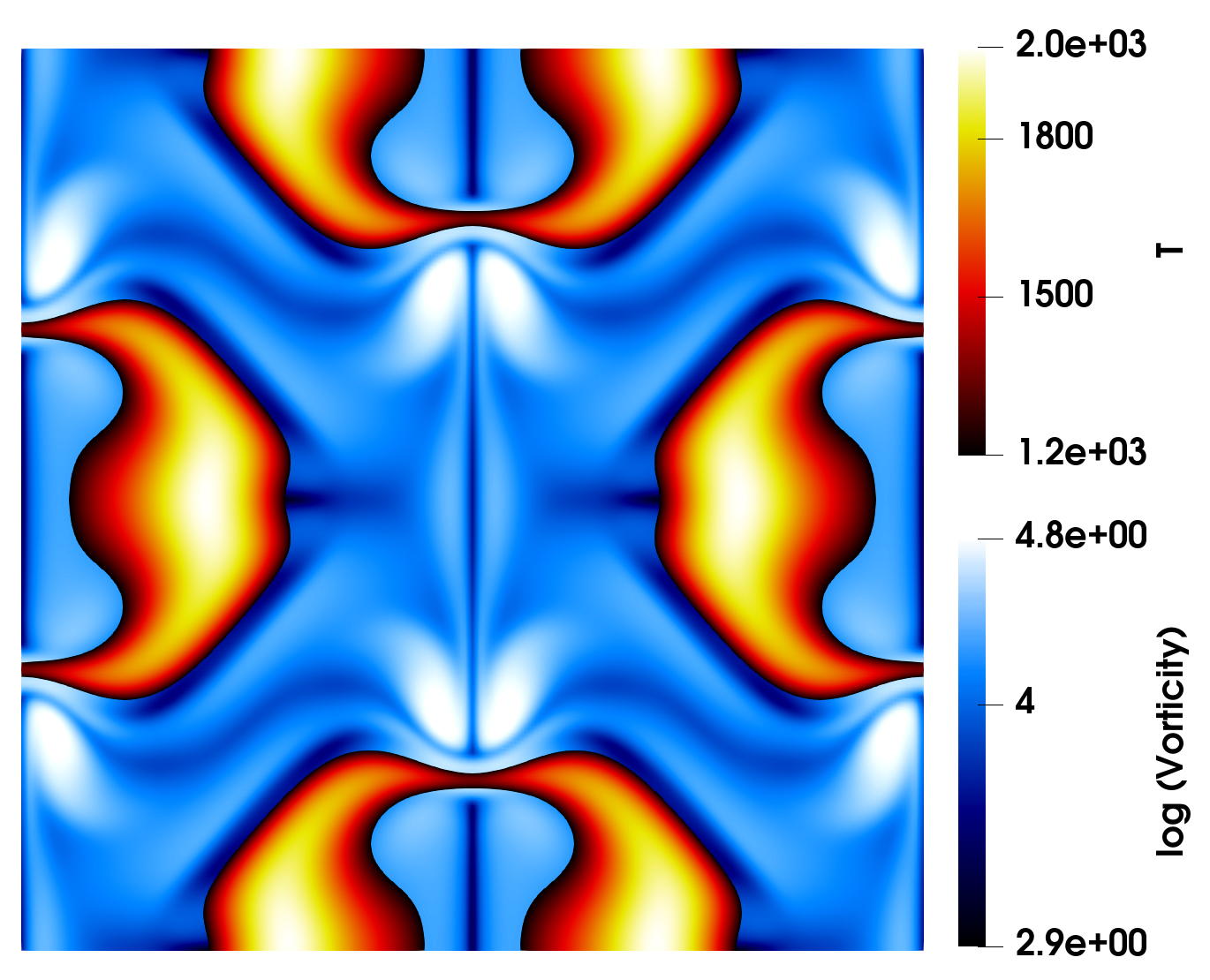}
		\end{minipage}
    }
    \subfigure[Case 1-UL]{
		\begin{minipage}[t]{0.45\linewidth}
			\centering
			\includegraphics[width=3in]{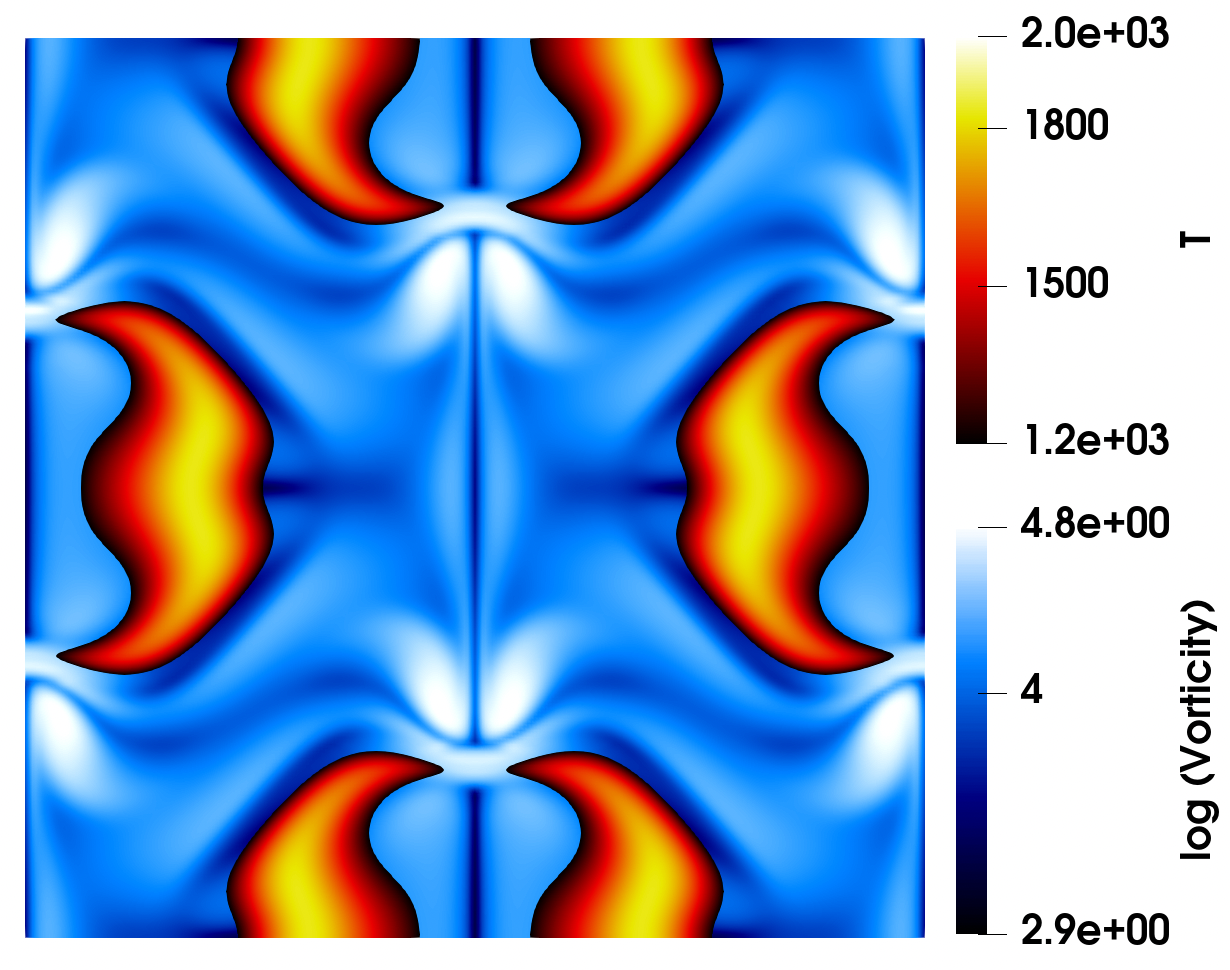}
		\end{minipage}
    }
    \subfigure[Case 2]{
		\begin{minipage}[t]{0.45\linewidth}
			\centering
			\includegraphics[width=3in]{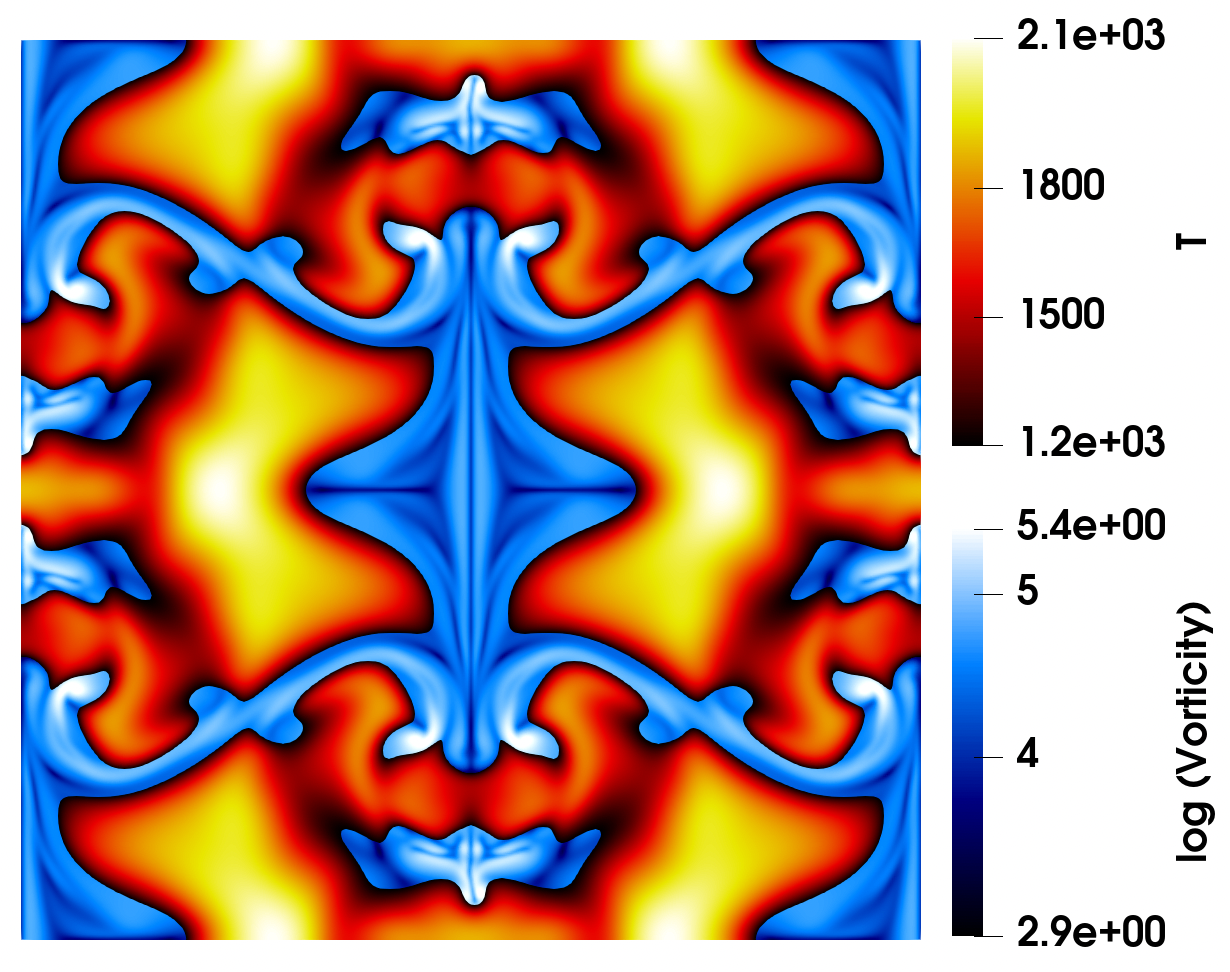}
		\end{minipage}
    }
    \subfigure[Case 2-UL]{
		\begin{minipage}[t]{0.45\linewidth}
			\centering
			\includegraphics[width=3in]{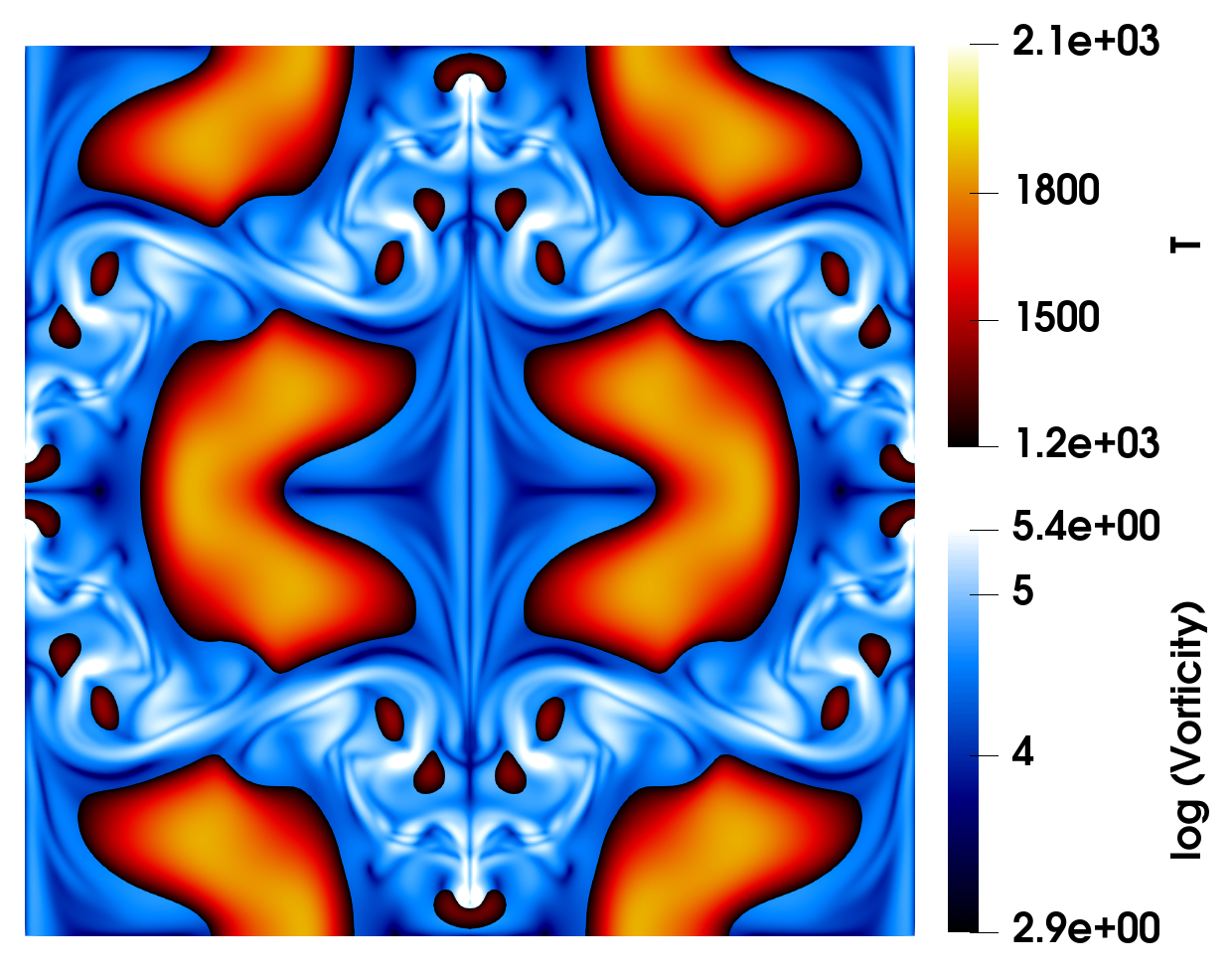}
		\end{minipage}
    }
    
    \subfigure[Case 3]{
		\begin{minipage}[t]{0.45\linewidth}
			\centering
			\includegraphics[width=3in]{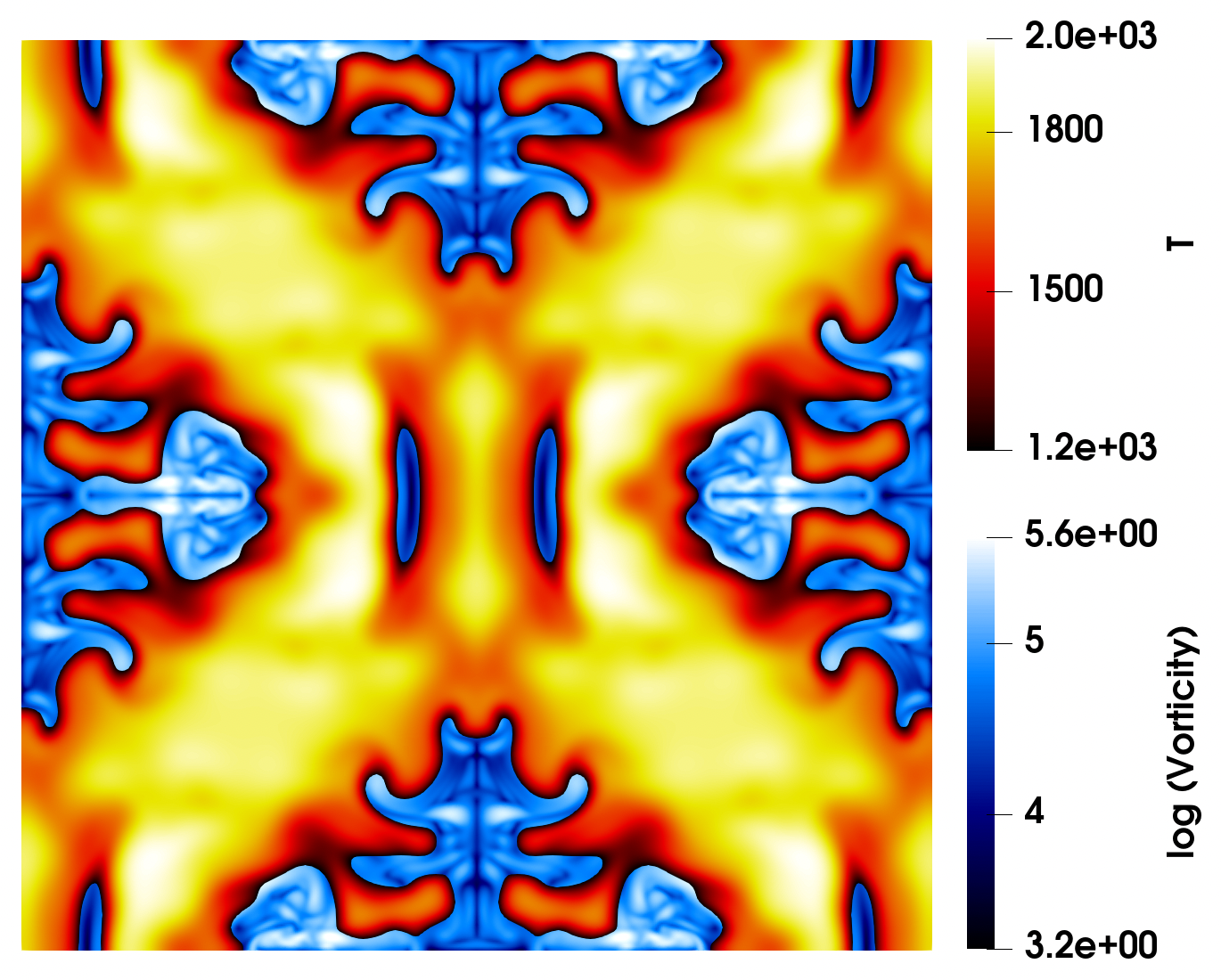}
		\end{minipage}
    }
    \subfigure[Case 3-UL]{
		\begin{minipage}[t]{0.45\linewidth}
			\centering
			\includegraphics[width=3in]{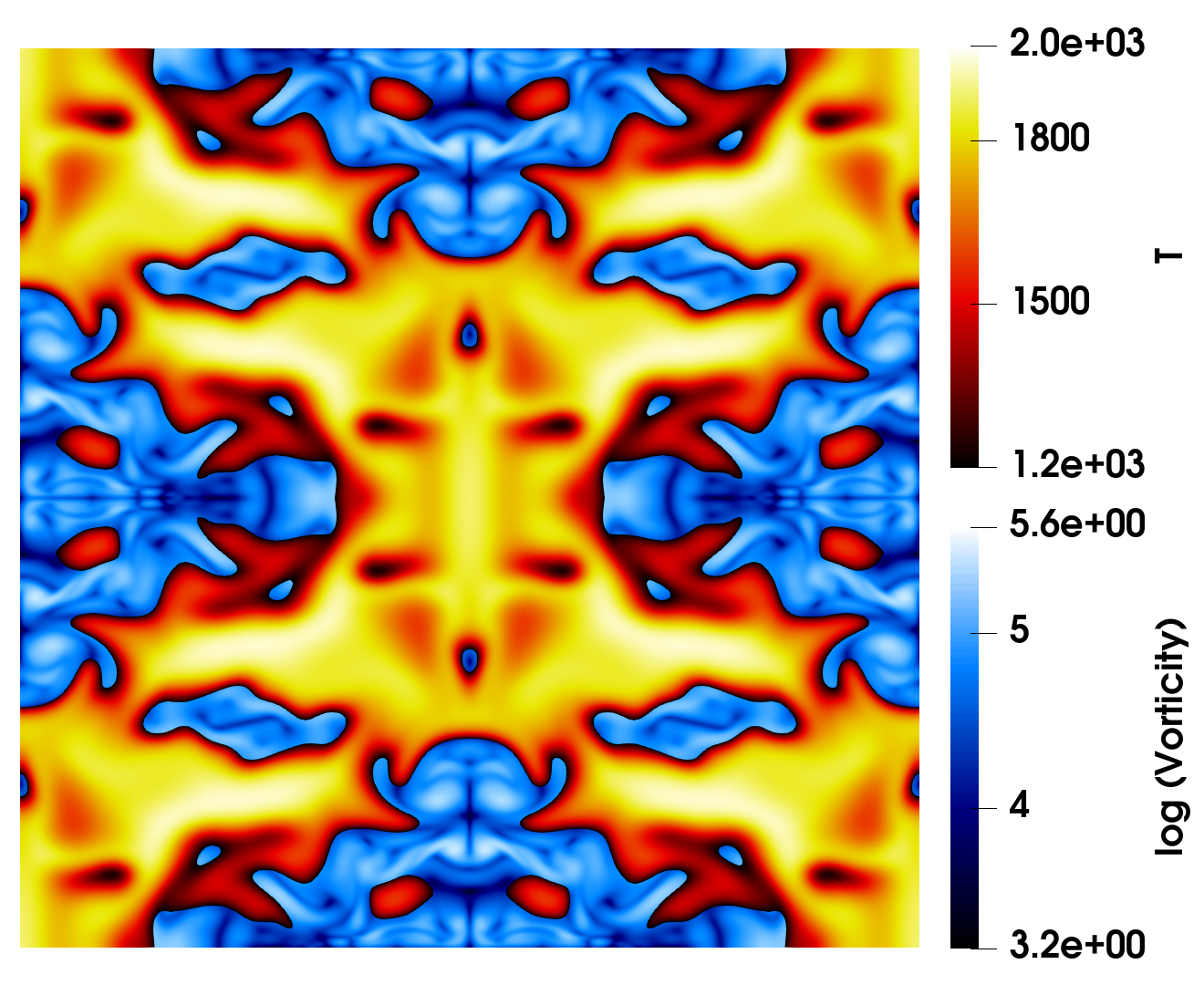}
		\end{minipage}
    }
    \centering
	\caption{Temperature and log (Vorticity) field in the $x = L/6, y-z$ plane at $t=6\tau_{chem}$ for the cases. In each line, cases with the same velocity magnitude and different diffusion model are displayed.}
	\label{fig8}
\end{figure*}

In this section, the effects of diffusion model (DM), namely the mixture-averaged and unity-Lewis number models, on the reacting TGV are analysed to investigate the transport and thermal effects of molecular diffusion. Also, since the molecular and turbulent mixing account for different proportions in cases with various Reynolds number, the evolution of flame and vortex helps to comprehensively assess DM effects under conditions with different turbulence intensity. 

To examine the thermal effects of DM on the flame structure and temperature, similar 2D-slices in the $x = L/6, y-z$ plane for the cases are shown in Fig. \ref{fig8}. In contrast to Fig.~\ref{fig6}, the subfigures in Fig.~\ref{fig8} are depicted for the same chemical reference time $t = 6\tau_{chem}$ rather than the peak time of enstrophy for each case. This is to keep the flame evolution state the same so as to effectively investigate the thermal and transport effects of DM. 

In Fig.~\ref{fig8}(a) and \ref{fig8}(b), the vortex field has not fully developed at $t = 6\tau_{chem}$ (corresponding to $t = 3.75\tau_f$ for Case 1 and Case 1-UL) indicating an early stage of evolution and weak turbulence intensity. In this situation, the transport and thermal effects of molecular diffusion play the prominent role in combustion. Compared with Fig. \ref{fig8}(b), it is demonstrated in Fig. \ref{fig8}(a) that the flame temperature is higher and the flame structure is thicker for Case 1 as the mixture-averaged model provides a more accurate simulation of the high diffusivity of hydrogen and its atom radicals leading to a stronger flame. In Fig. \ref{fig8}(c) and Fig. \ref{fig8}(d), vortex tubes and filaments can be observed at $t = 6\tau_{chem}$ (corresponding to $t = 7.5\tau_f$ for Case 2 and Case 2-UL) where the combustion region is apparently broader and the flame temperature is higher for Case 2. Fig. \ref{fig8}(e) and Fig. \ref{fig8}(f) show that the turbulent field has fully developed at $t = 6\tau_{chem}$ (corresponding to $t = 15\tau_f$ for Case 3 and Case 3-UL). As a result, turbulent intensity provides adequate mixing of the species which is a dominant process rather than molecular mixing. Under the effects of turbulent mixing, combustion fills up the computing domain where the distinctions of the flame strength between the 2 cases diminish relatively. 

\begin{figure*}[htbp]
    \centering
    \subfigure[Case 1/1-UL]{
		\begin{minipage}[t]{\linewidth}
			\centering
			\includegraphics[width=0.3\linewidth]{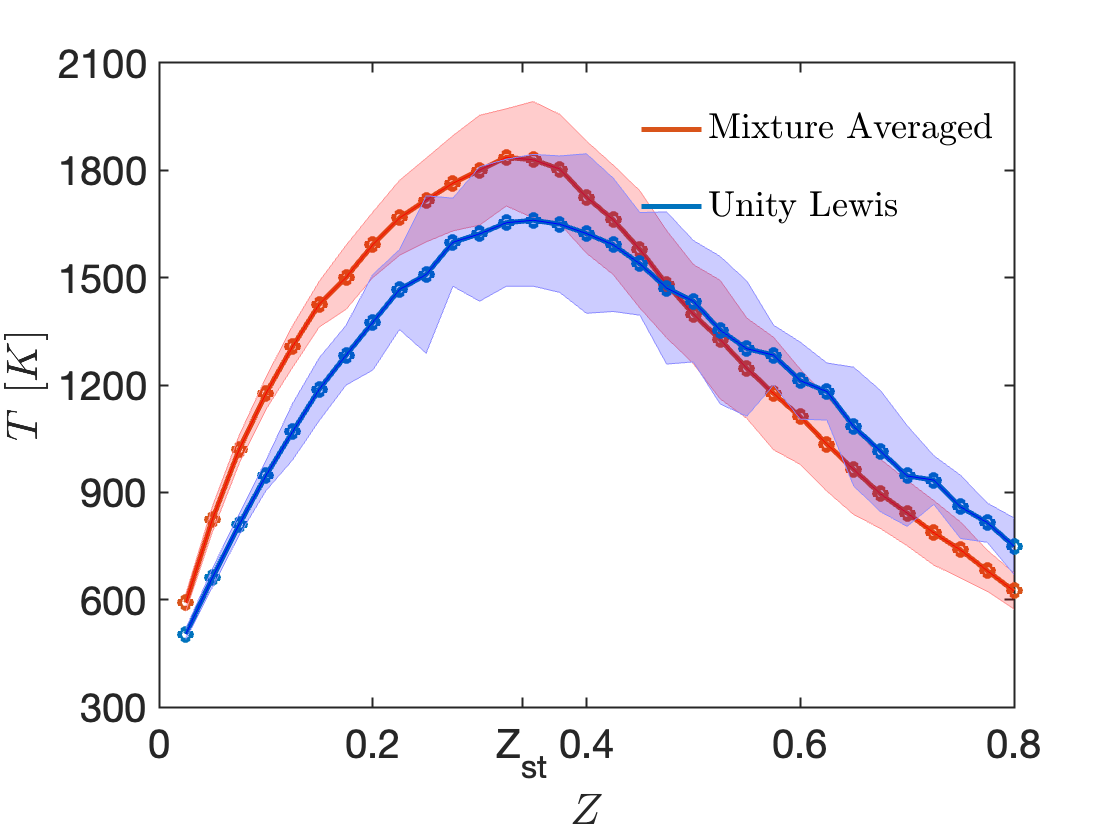}
            \includegraphics[width=0.3\linewidth]{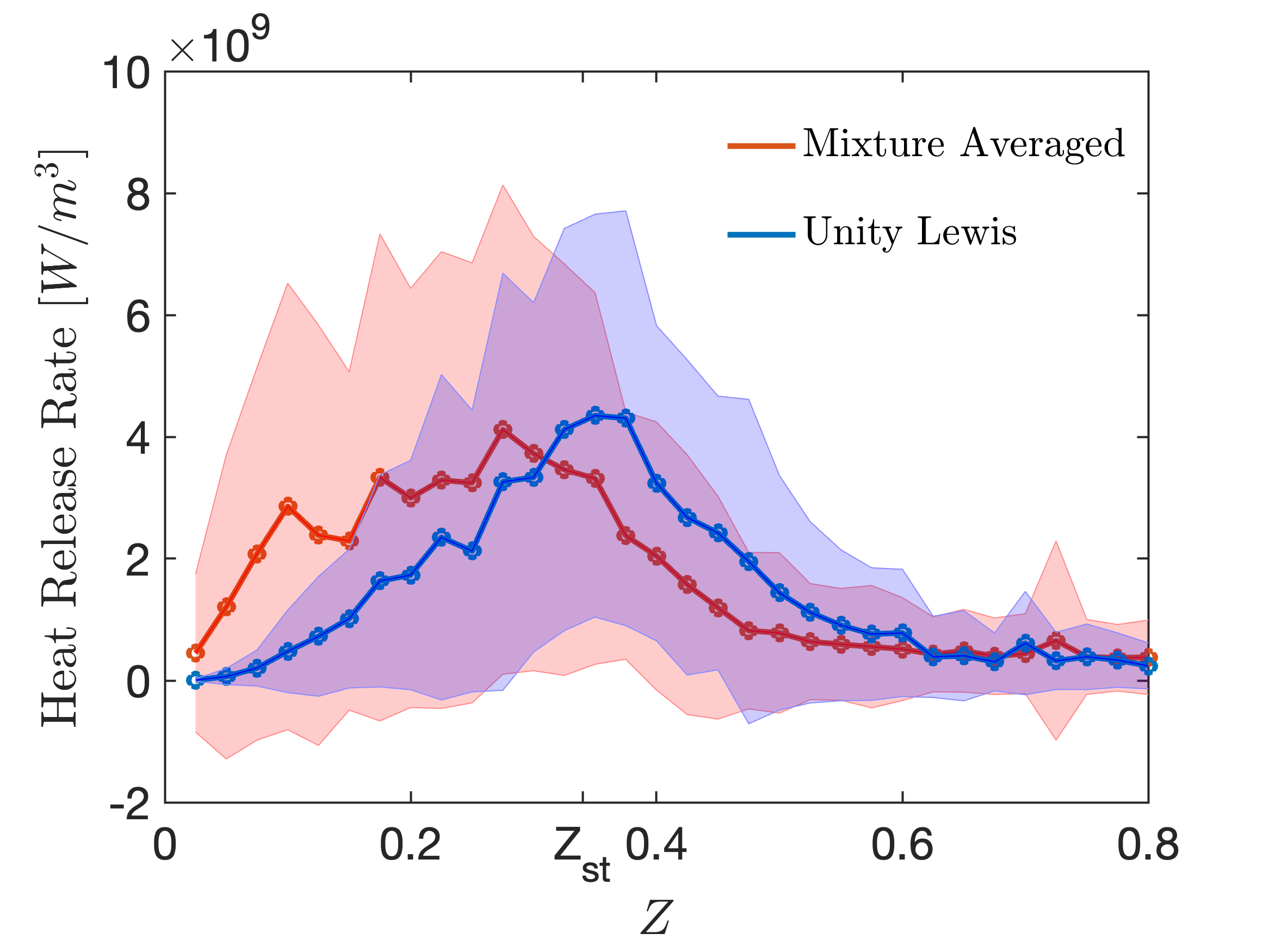}
            \includegraphics[width=0.3\linewidth]{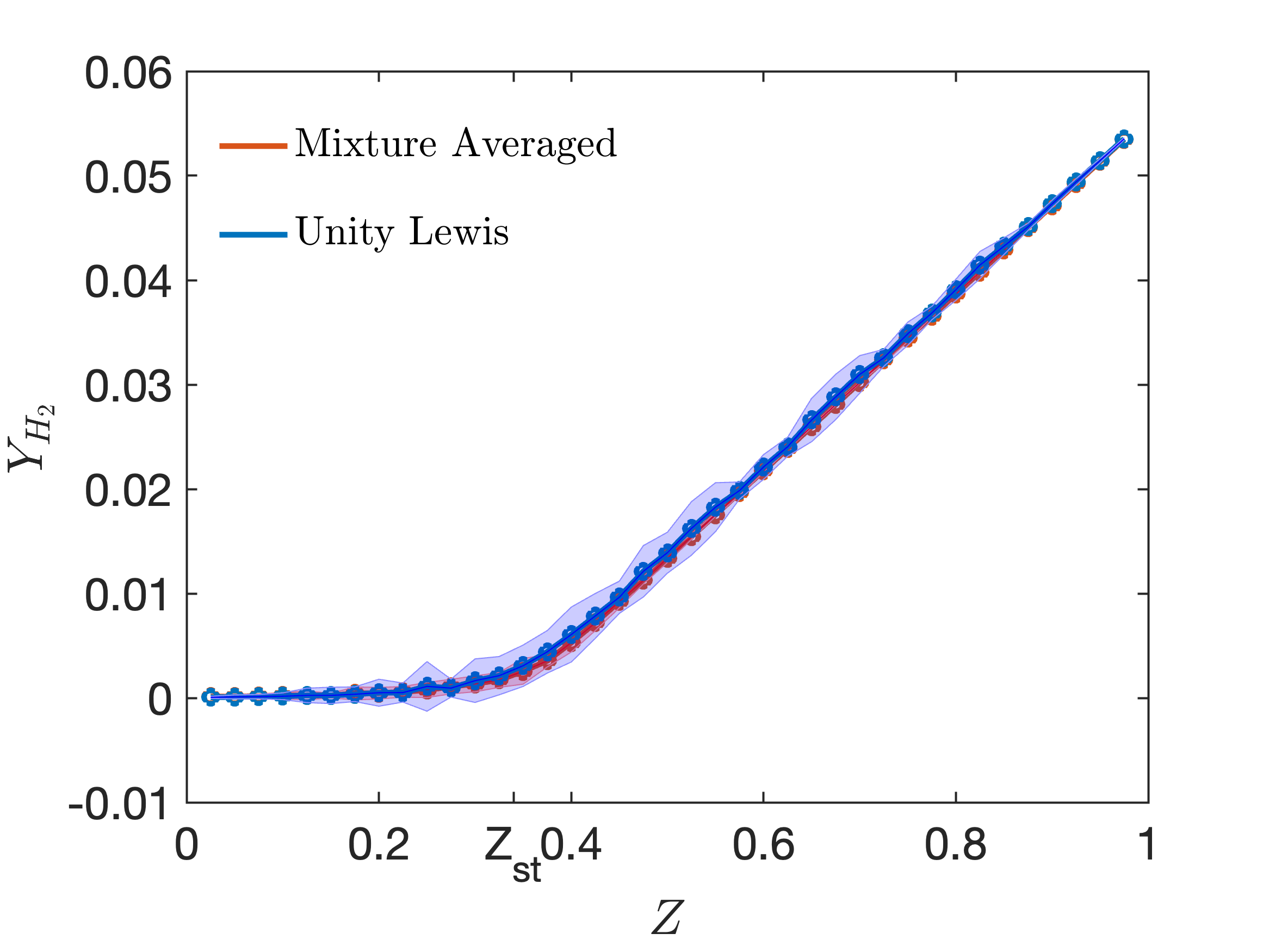}
		\end{minipage}
    }
    \subfigure[Case 2/2-UL]{
		\begin{minipage}[t]{\linewidth}
			\centering
			\includegraphics[width=0.3\linewidth]{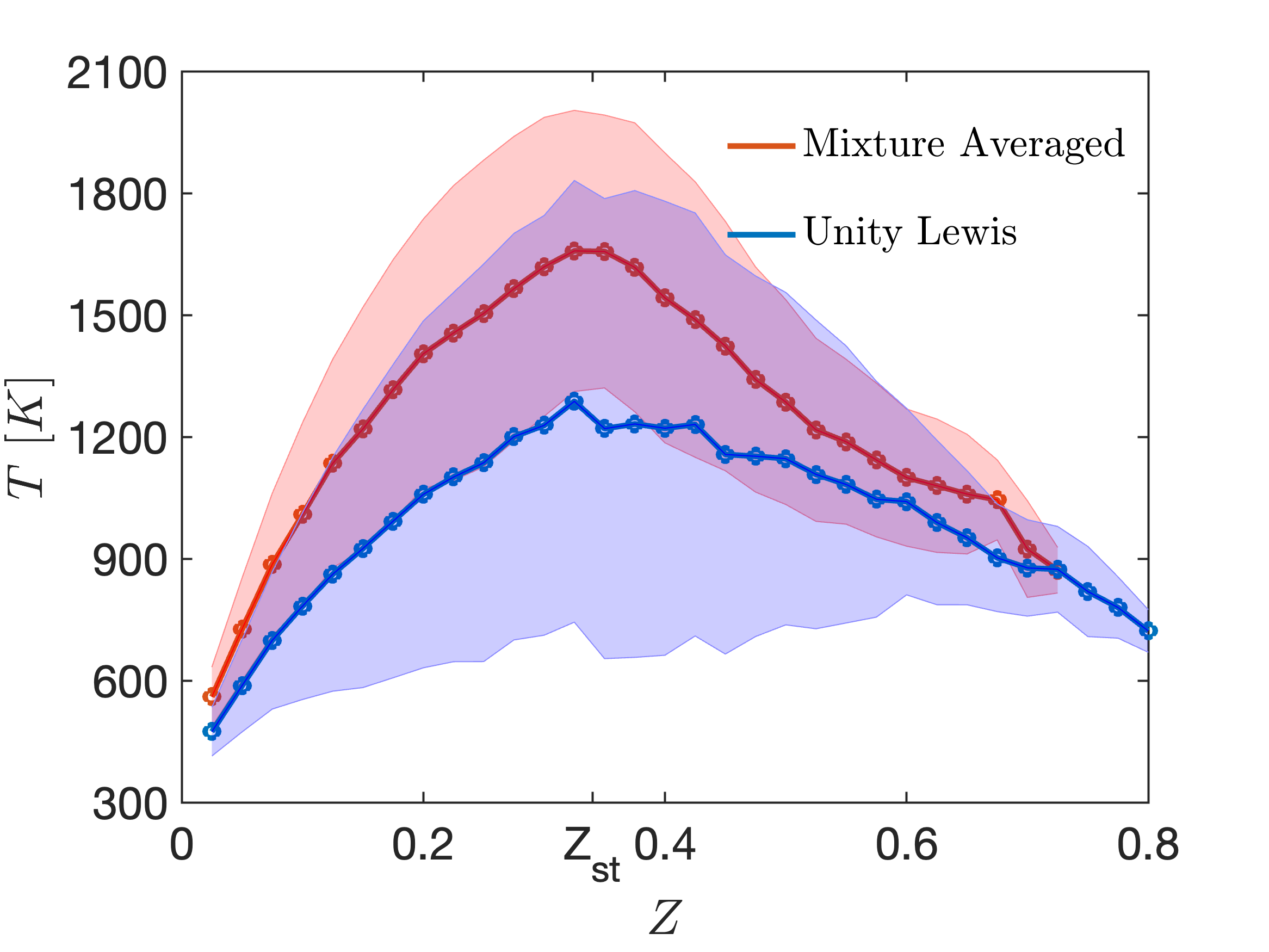}
            \includegraphics[width=0.3\linewidth]{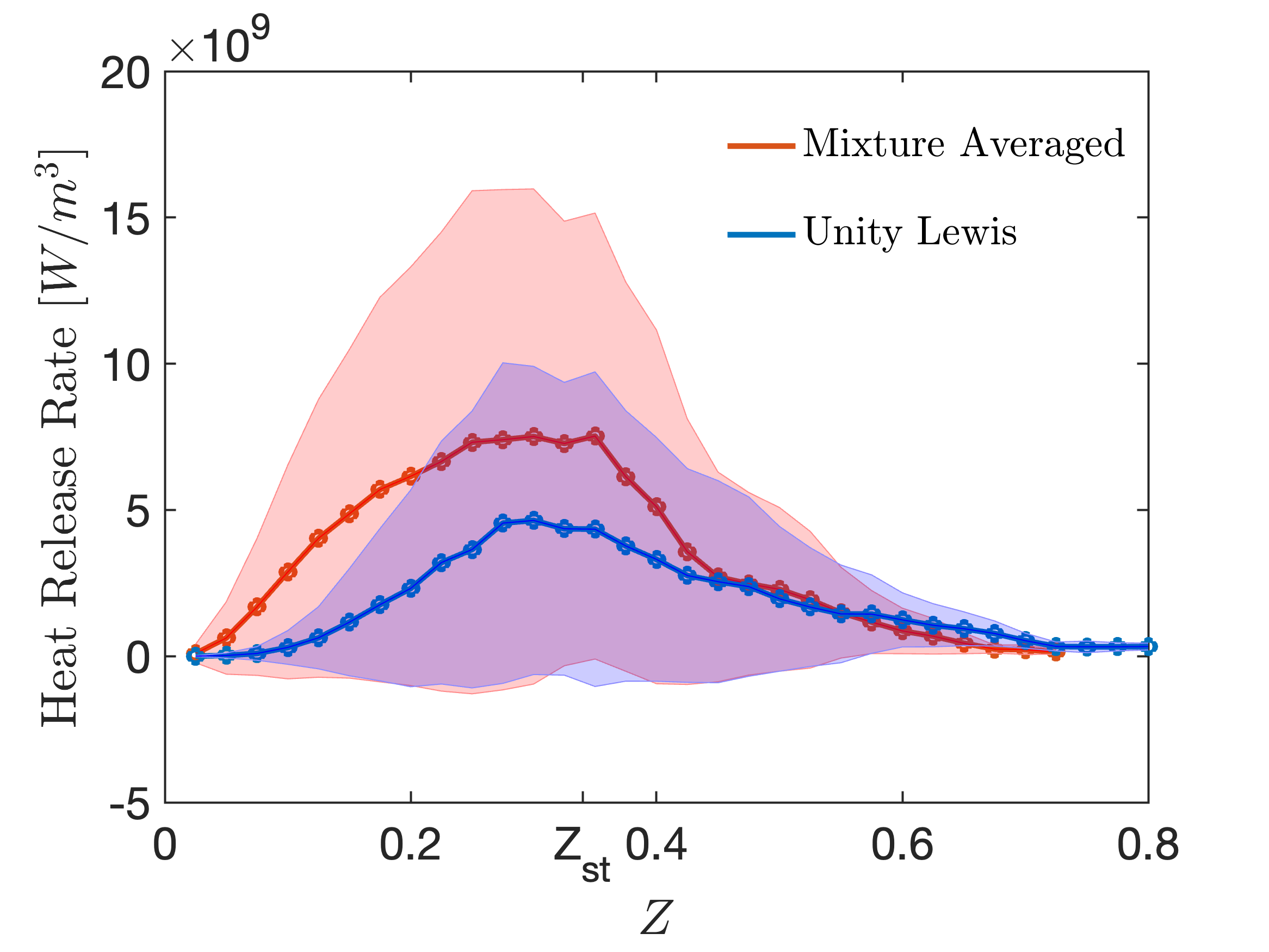}
            \includegraphics[width=0.3\linewidth]{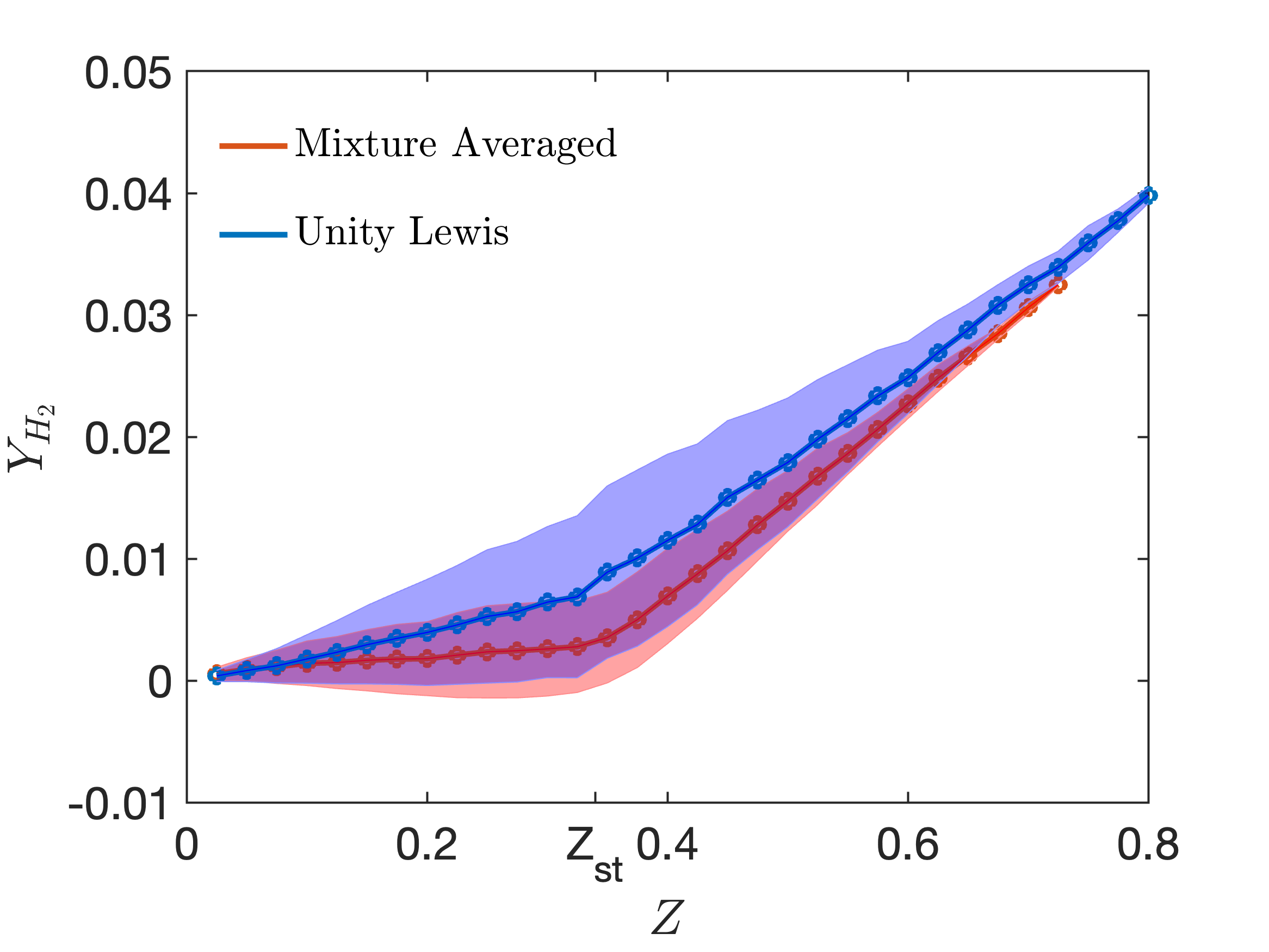}
		\end{minipage}
    }
    \subfigure[Case 3/3-UL]{
		\begin{minipage}[t]{\linewidth}
			\centering
			\includegraphics[width=0.3\linewidth]{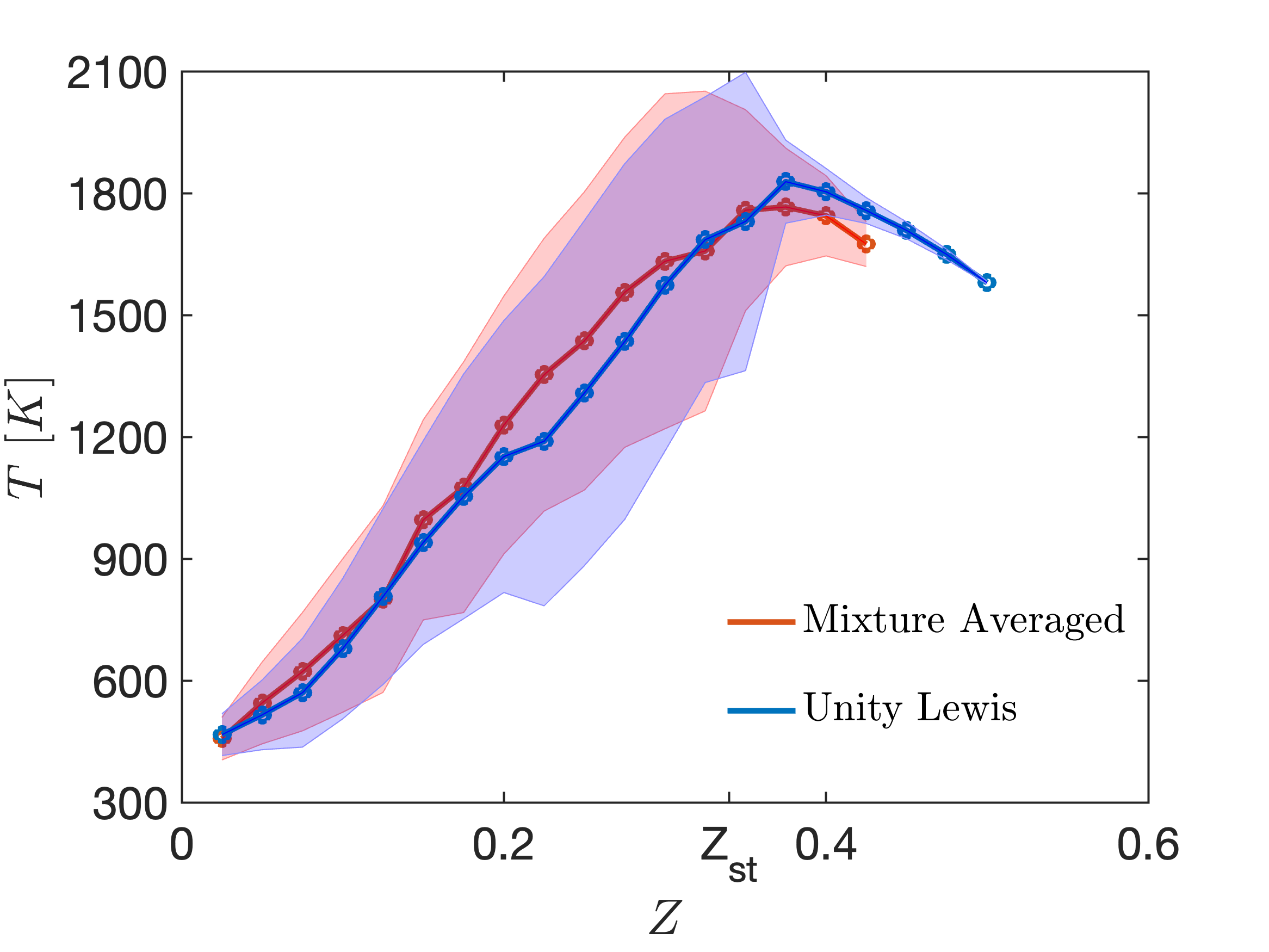}
            \includegraphics[width=0.3\linewidth]{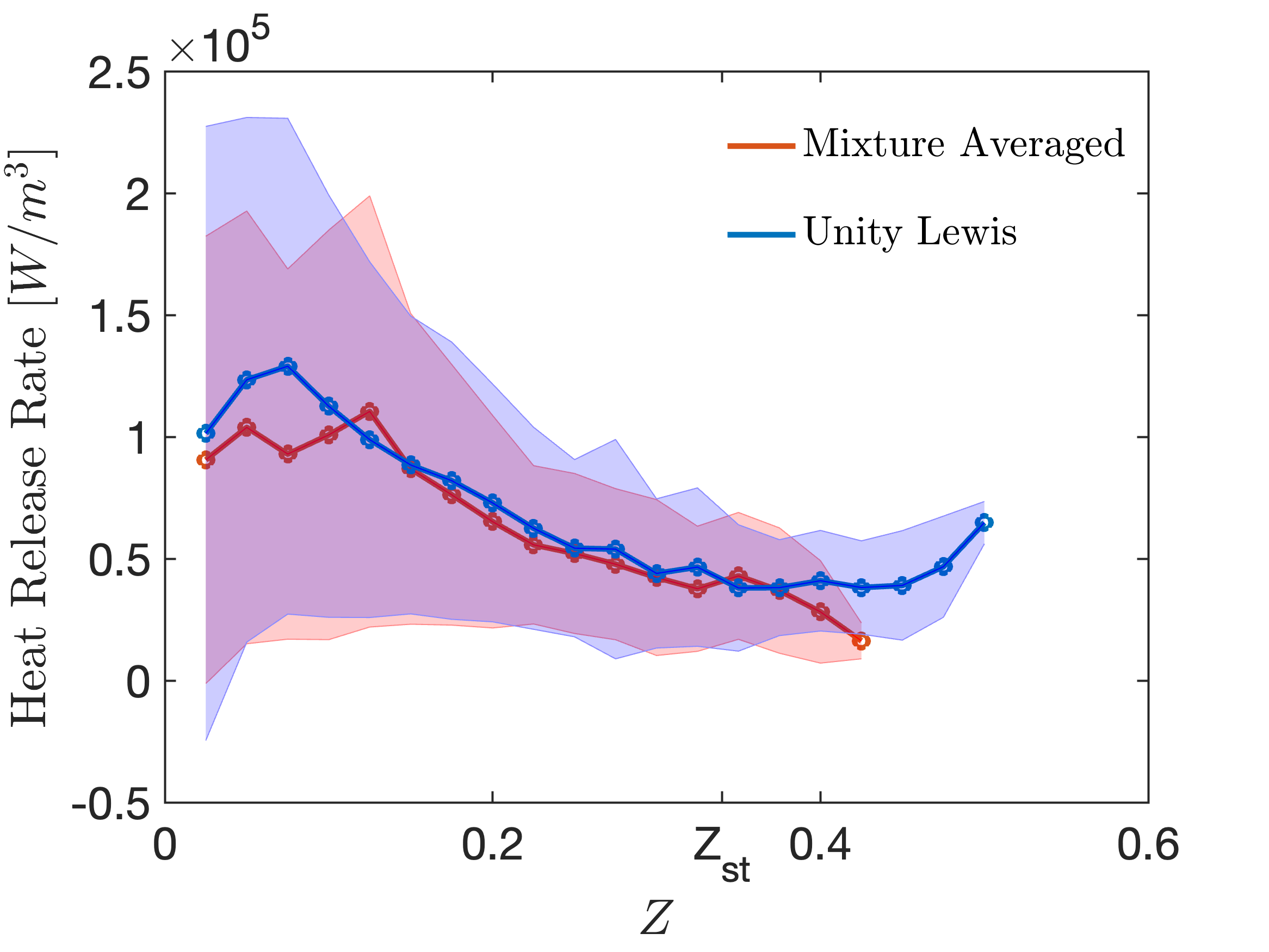}
            \includegraphics[width=0.3\linewidth]{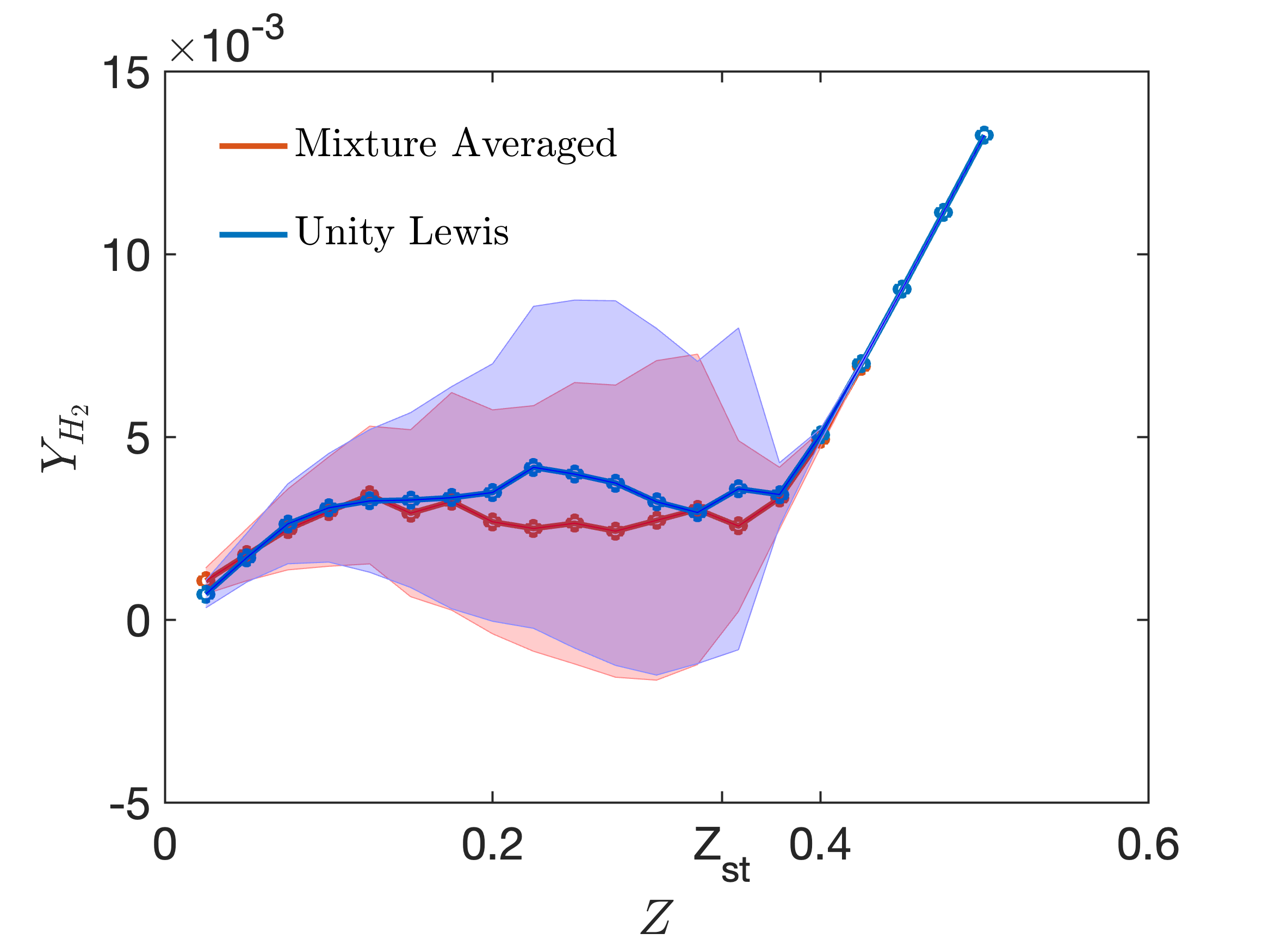}
		\end{minipage}
    }
    \centering
	\caption{Conditional means and standard deviations of temperature, heat release rate and mass fraction of H$_2$ as functions of mixture fraction $Z$ at $t = 6\tau_{chem}$. 
    }
	\label{fig9}
\end{figure*}

For a more quantitative analysis of the DM effects, the conditional mean profiles and standard deviations of typical quantities such as temperature, heat release rate and mass fraction of H$_2$ as functions of mixture fraction $Z$ at $t = 6\tau_{chem}$ are compared and shown in Fig. \ref{fig9}. In Fig. \ref{fig9}(a), it is seen that the conditional mean value of temperature is higher for Case 1, which is consistent with the field plot shown in Fig.~\ref{fig8}(a) and (b). The standard deviation of the conditional temperature and $Y_{H_2}$ are quite small because the combustion at this low Reynolds number condition concentrates in specific areas at the early stage of vortex evolution (see Fig.~\ref{fig3}). It is seen in Fig.~\ref{fig9}(b) that the variation of the conditional mean value of temperature increases and the corresponding standard deviation area becomes wider. This is because the vortex-induced motions accelerate the mixing of the species resulting in a broader range of thermochemical status with varying mixture fraction in the domain. As a fully evolved turbulence field has formed at $t = 6\tau_{chem}$ for Case 3 and Case 3-UL, Fig.~\ref{fig9}(c) demonstrates that the effects of the diffusion model are reduced due to the significant turbulence intensity dominating the diffusion processes. 

\begin{figure*}[h!]
    \centering
    \subfigure[]{
		\begin{minipage}[t]{0.45\linewidth}
			\centering
			\includegraphics[width=\linewidth]{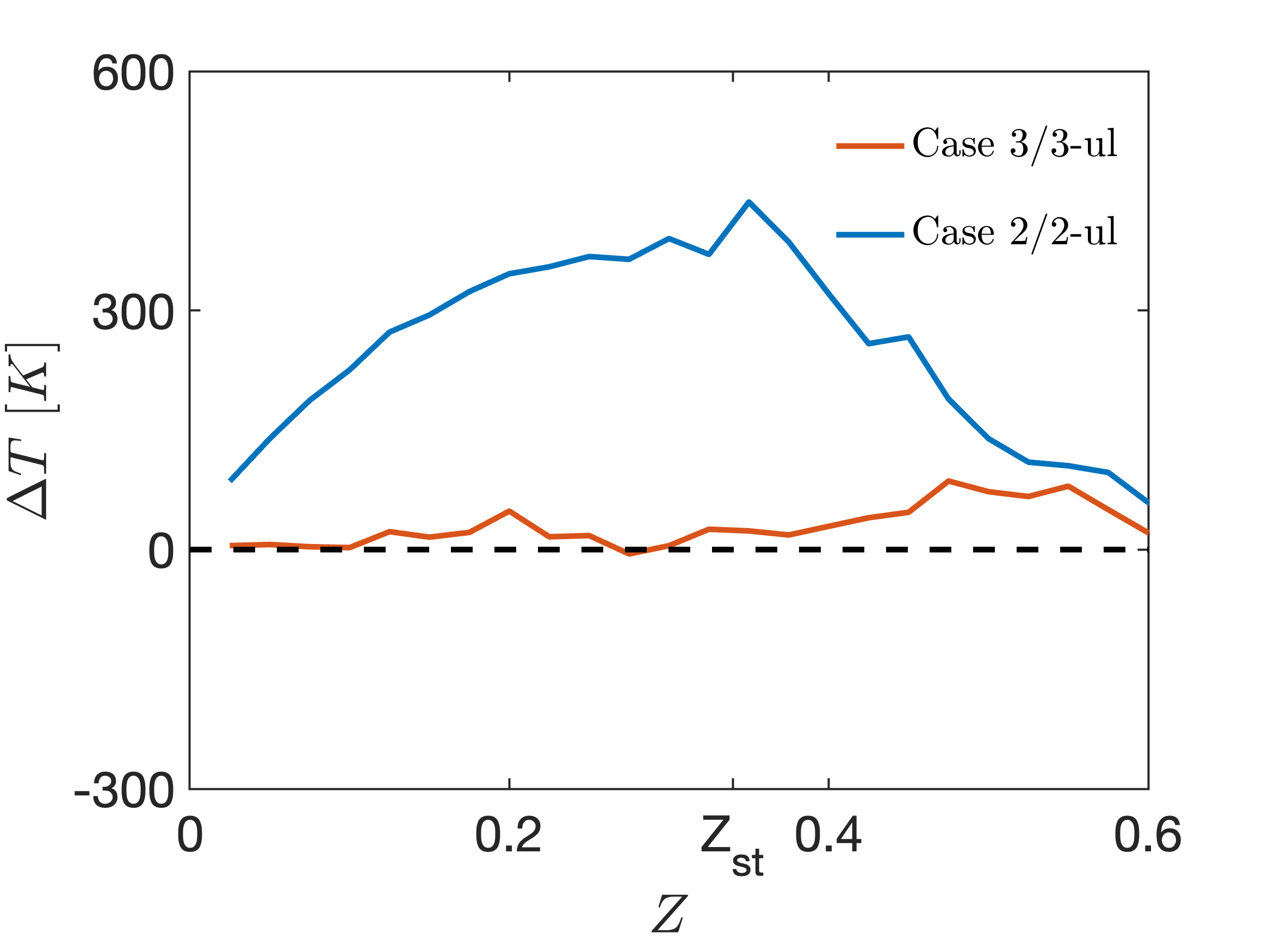}
		\end{minipage}
    }
    \subfigure[]{
		\begin{minipage}[t]{0.45\linewidth}
			\centering
			\includegraphics[width=\linewidth]{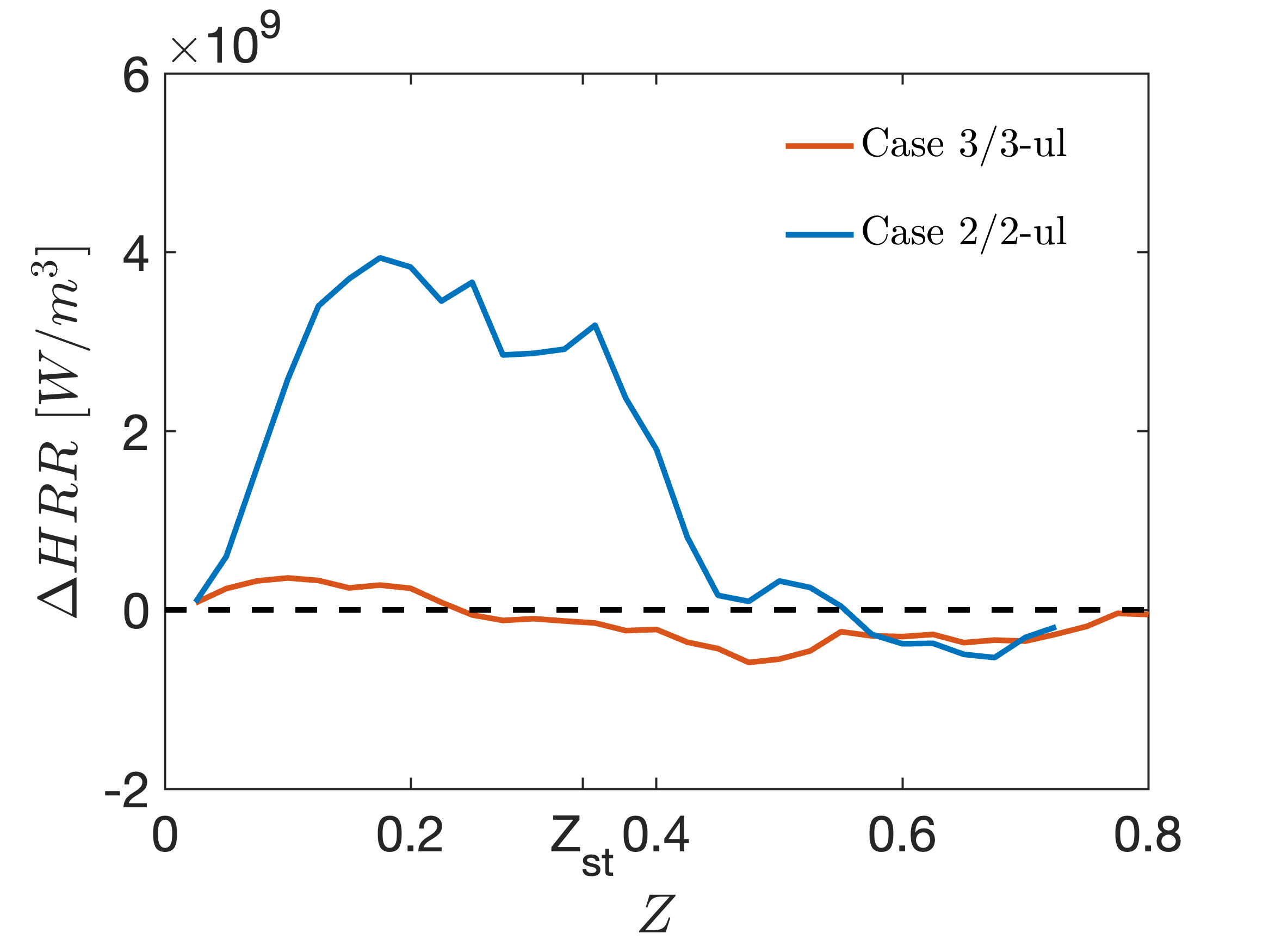}
		\end{minipage}
    }
    \centering
	\caption{Conditional mean differences of temperature and heat release rate between Case 2 (3) and Case 2-UL (3-UL) at $t = 8\tau_f$ for each case.}
	\label{fig10}
\end{figure*}

To further investigate the turbulence effects on the DM, the conditional mean differences of temperature and heat release rate between the Cases and Cases-UL are calculated and shown in Fig.~\ref{fig10}. As the developments of the vortex field in Case 1 and Case 1-UL are relatively slow, the mean differences between Case 2 (3) and Case 2-UL (3-UL) are depicted at an identical flow reference time $t = 8\tau_f$. This is to keep the same state of vortex evolution for the cases shown. It can be observed in the figure that the mean differences of temperature and HRR between Case 3 and Case 3-UL are much smaller than that between Case 2 and Case 2-UL, which further confirms that turbulence can significantly reduce the effects of diffusion models as speculated earlier from Fig.~\ref{fig9}.

\subsection{Impact of vortex length scale}

The above discussion analysed the flame/TGV interactions under the effects of Reynolds number resulting from a varying velocity magnitude. In this subsection, the impact of vortex length scale which can also contribute to the Reynolds number variation is investigated (see Fig.~\ref{fig1}). Due to the distinct characteristics of the present TGV-flame setup, the flow time scale $\tau_f$ varies with both velocity magnitude and length scale, while the chemical time scale $\tau_{chem}$ is kept constant. To further investigate the TGV-flame cases via controlling variables and complete the dataset of the simulations to cover a broader area in the regime diagram in Fig.~\ref{fig1}, Case 4/4-UL with $u_0 = 25 $~m/s, $L_0 = 2$~mm is implemented. This setup has the same $\tau_f$ corresponding to the same $Da$ as Case 1 and the same $Re$ as Case 3. 

\begin{figure*}[htbp]
    \centering
    \subfigure[Case 4]{
		\begin{minipage}[t]{0.3\linewidth}
			\centering
			\includegraphics[width=\linewidth]{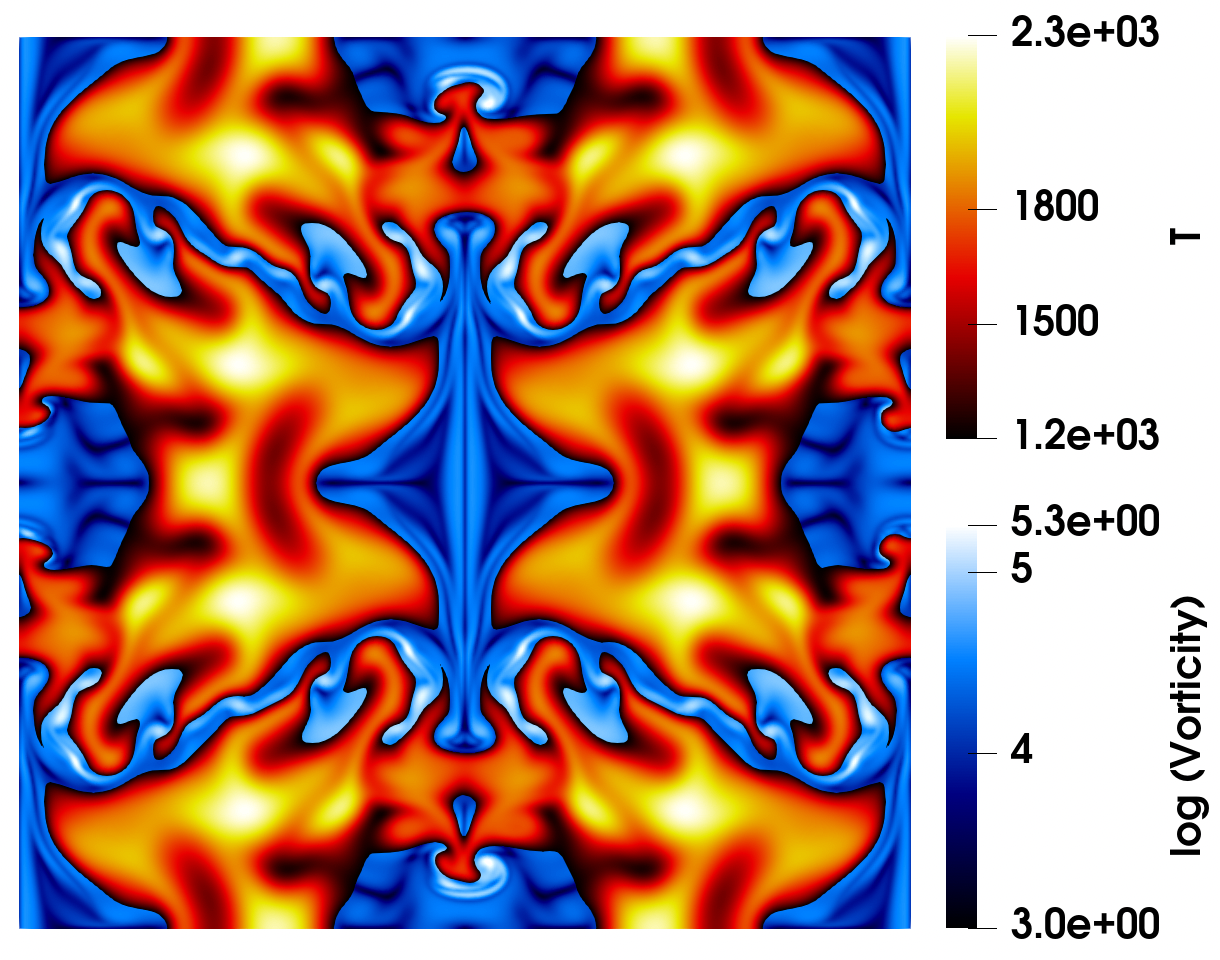}
		\end{minipage}
    }
    \subfigure[Case 1]{
		\begin{minipage}[t]{0.3\linewidth}
			\centering
			\includegraphics[width=\linewidth]{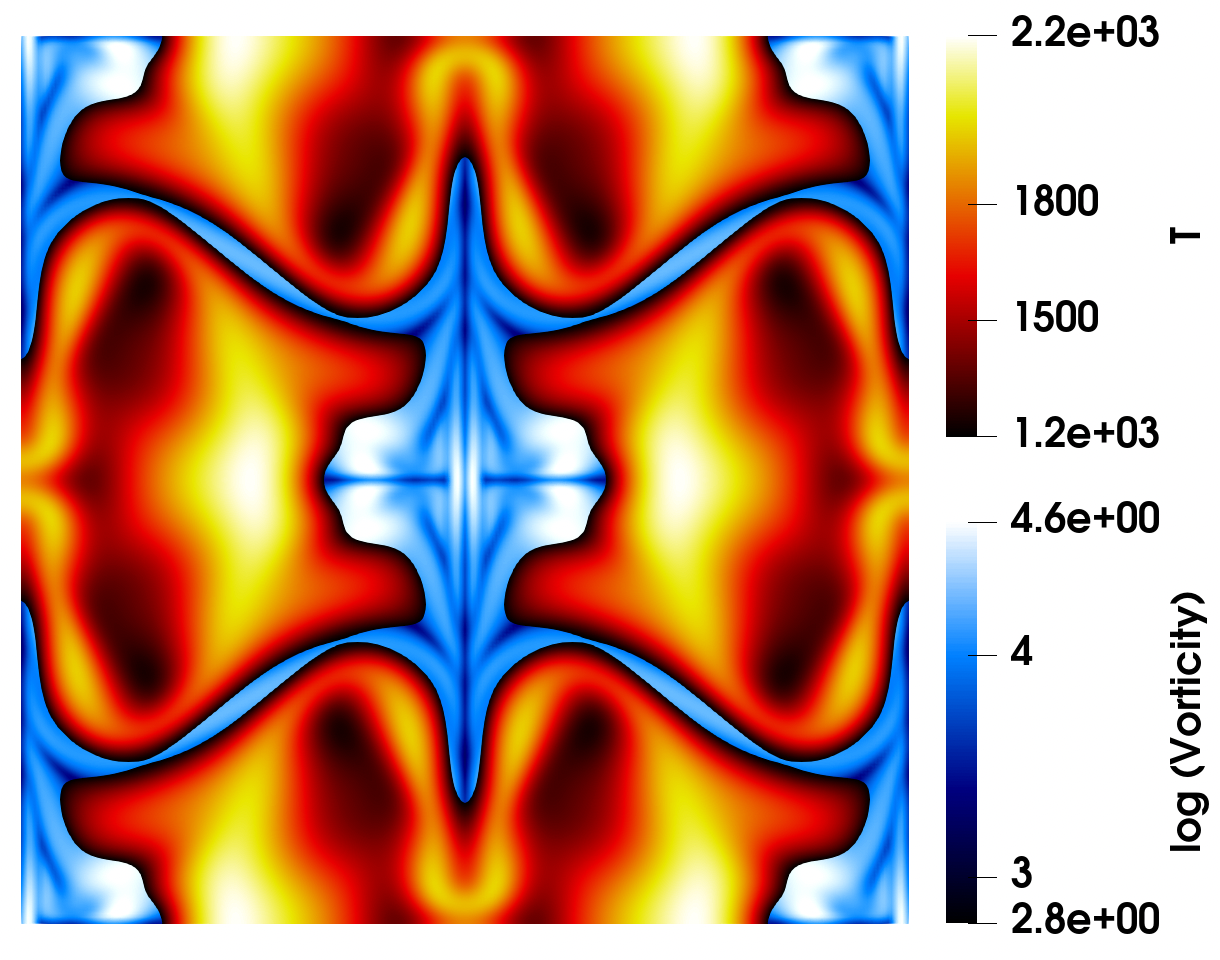}
		\end{minipage}
    }
    \subfigure[Case 3]{
		\begin{minipage}[t]{0.3\linewidth}
			\centering
			\includegraphics[width=\linewidth]{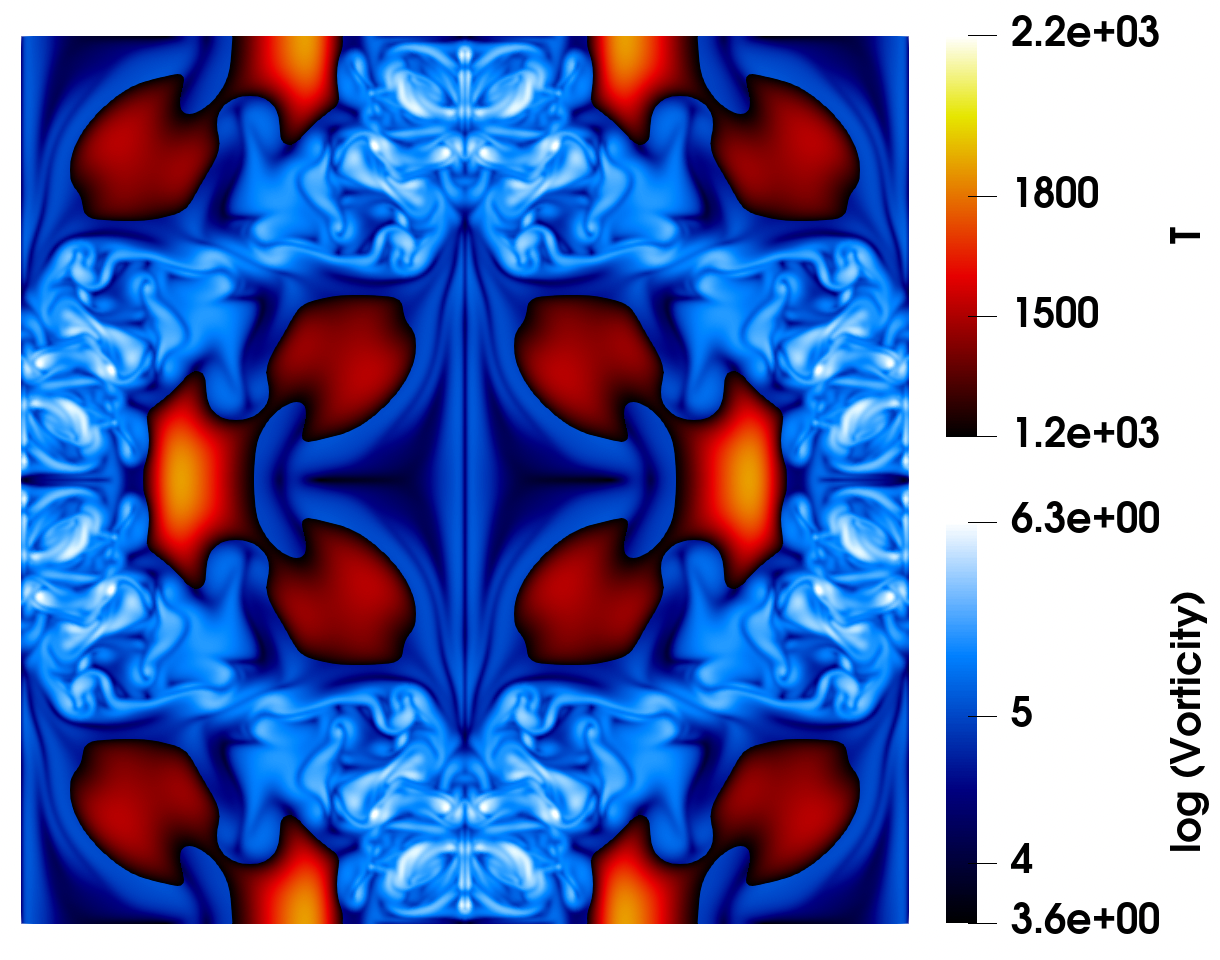}
		\end{minipage}
    }
    \centering
	\caption{Temperature and log (Vorticity) field in the $x = L/6, y-z$ plane at $t = 8\tau_f$ for Case 4, Case 1 and Case 3.}
	\label{fig11}
\end{figure*}

\begin{figure}[htbp]
\centering
\includegraphics[width=0.7\linewidth]{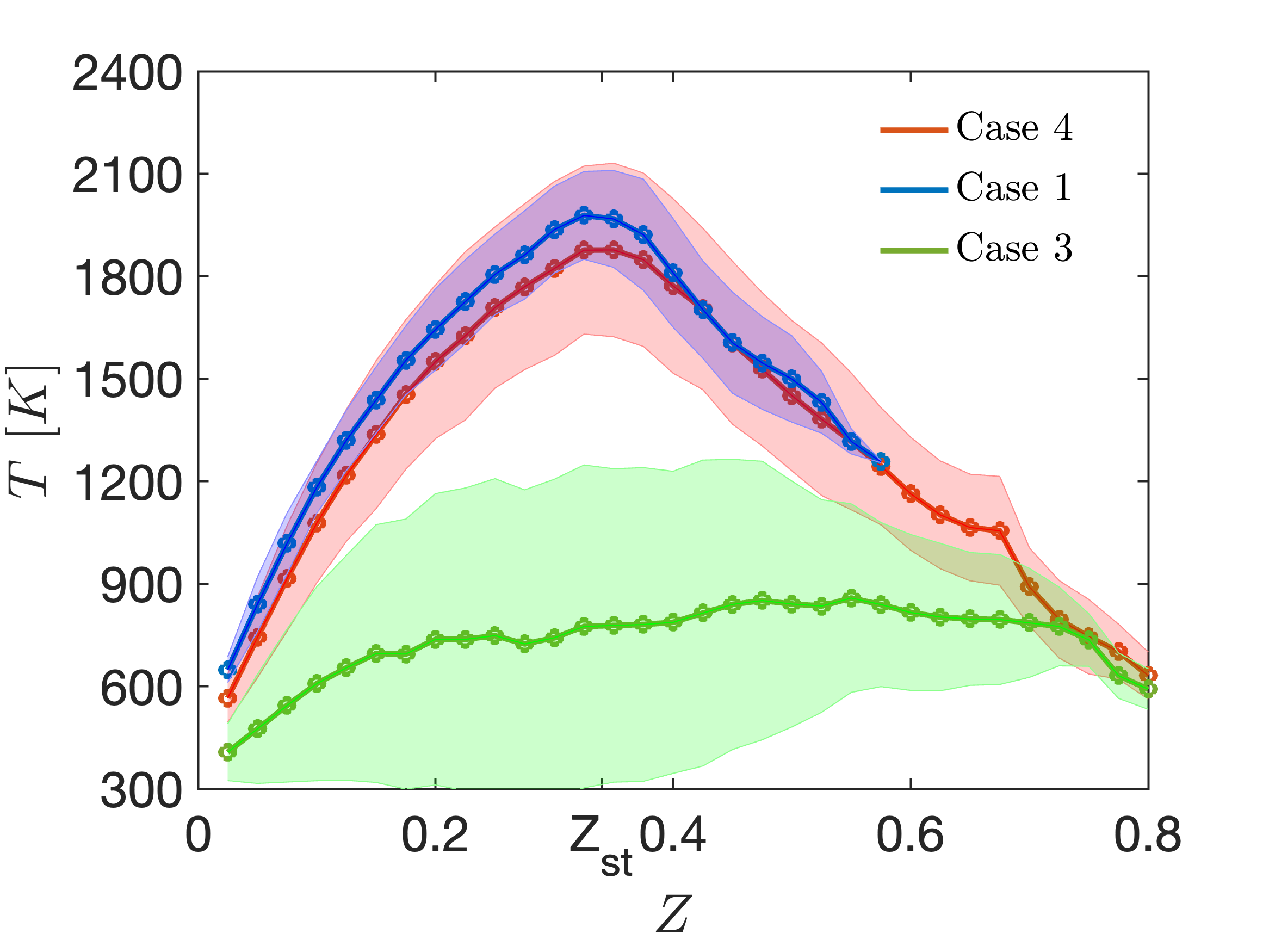}
\caption{Conditional means and standard deviations of temperature on mixture fraction $Z$ at $t = 8\tau_f$ for Case 4, Case 1 and Case 3.}
\label{fig12}
\end{figure}

\begin{figure}[htbp]
\centering
\includegraphics[width=0.7\linewidth]{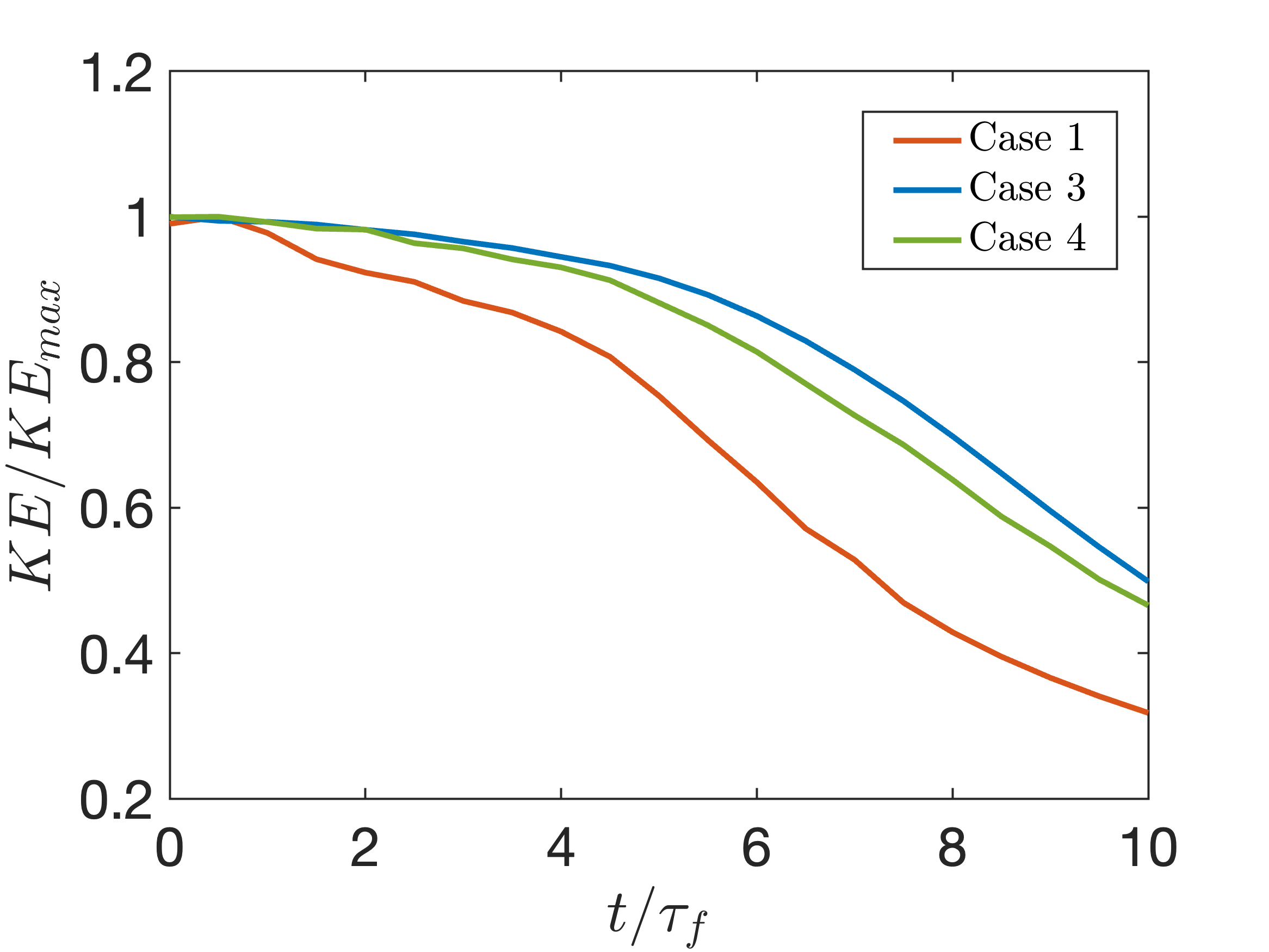}
\caption{Temporal evolution of the volume-averaged kinetic energy for Case 4, Case 1 and Case 3.}
\label{fig13}
\end{figure}

Figure~\ref{fig11} shows the temperature and vorticity field in the $x = L/6, y-z$ plane at $t = 8\tau_f$ for Case 4, Case 1 and Case 3. The conditional means and standard deviations of temperature on mixture fraction at the corresponding time for the same cases are plotted in Fig.~\ref{fig12}. As Case 1 and Case 4 have identical values of $\tau_f$ and $\tau_{chem}$, the vortical evolution and combustion develop at the relative pace for the 2 cases. Combining Figs.~\ref{fig11}(a) and \ref{fig11}(b) having the same $Da$, it is revealed that the significant turbulence intensity leads to significantly stronger flame wrinkling in Case 4. As seen in Fig.~\ref{fig12}, the standard deviation range of Case 4 is wider than Case 1 as one would expect. Considering Case 4 and Case 3 with the same $Re$, it is demonstrated in Fig.~\ref{fig11}(a) and \ref{fig11}(c) that the vortex field is more intense in Case 4 as the combustion is still at an early stage at $t=8\tau_f$ for Case 3 (i.e. due to smaller $Da$) as shown in Fig.~\ref{fig12}. As a result, a relatively weaker suppression effect of the flame on turbulence can be observed in Fig.~\ref{fig11} for Case 3 compared to Case 4. 

In order to investigate the combustion effects on turbulence intensity, the temporal evolution of the normalised volume-averaged kinetic energy for the cases scaled in the flow reference time $\tau_f$ are shown in Fig.~\ref{fig13}. As seen in the figure, the kinetic energy of Case 1 dissipates rapidly because of its weak vortical motions. Comparing Case 4 with Case 1, the dissipation of the kinetic energy is slower in Case 4 which confirms the speculation in Subsection A that the relative delay of combustion leads to a relative longer evolution time of turbulence. The faster decay of kinetic energy in Case 4 than in Case 3 further demonstrates the flame suppression effect on the turbulence seen earlier in Fig.~\ref{fig11}.

\section{Summary and Conclusions}
In this work, the interactions between a hydrogen diffusion flame and the 3D Taylor-Green Vortex are investigated and analysed by performing several DNS cases considering different Reynolds numbers and molecular diffusion models. The Reynolds effects are investigated by varying both the velocity magnitude and length scale of the TGV. The evolution of the flow field which is driven by both flame-induced and vortex-induced motions is studied in detail. The results show that the lumpy vortices are first rolled up into thinner vortex tubes in the early evolving stage of the TGV for all cases. Reynolds number and flame-induced effects do not play a significantly role in this stage. Subsequently, vortices start to dissipate subsequently in cases with smaller velocity magnitude, while the consistent stretching, splitting and twisting of vortex tubes can be observed in cases with higher turbulence intensity. 

Due to the distinct characteristics of the TGV, the vortex field interacts with the combustion process in a complex manner. With the increase of velocity magnitude, the significant turbulence intensity provides efficient mixing and kinetic energy for combustion. Meanwhile, the flow time scale decreases compared to the chemical time scale, resulting in that the development of chemical reaction, as well as its suppression effects, are relatively delayed leading to a longer evolution time of turbulence.


Because of the high diffusivity of hydrogen, the influence of diffusion model is investigated by comparing mixture-averaged and unity-Lewis number models. The results show that the molecular diffusion effects dominate the mixing process in the cases with weak vortices, resulting in a thicker flame with higher temperature using the mixture-averaged diffusion model. With the development of vortices and increasing turbulence intensity, the vortex-induced motions accelerate the mixing and becomes dominant over the molecular diffusion. 

It is noted that the current work a first attempt using DNS in exploring the various dynamics in the increasingly popular benchmark case of a hydrogen diffusion flame embedded in the Taylor-Green Vortex. To investigate the complex dynamics such as the vortex-induced quenching and relaminarization, further studies are required in future works.

\section*{Appendix A. Validation of simulation}
As reported in Section 2A, the DNS cases are conducted using the in-house finite-difference code, Advanced flow Simulator for Turbulence Research (ASTR). To validate our approach and solver, step 4 in the benchmark proposed by Abdelsamie et al.\cite{abdelsamie_taylorgreen_2021} is followed referring to a 3-D reacting mixture. With the basic configuration introduced in Section 2B, the computing domain of the case is a $[0,L]^3$ cubic with $L = 2\pi L_0$, where $L_0 = 1$~mm. The velocity magnitude is settled as $u_0 = 4$~m/s resulting in $Re \approx 250$ and $\tau_f = 0.25$~ms. A 12 step skeletal hydrogen–air mechanism\cite{BOIVIN2011517} is considered for the case. 

In the following, we compare the results obtained with ASTR on a $256^3$ grid, with the results reported by Abdelsamie et al.\cite{abdelsamie_taylorgreen_2021} with NEK5000. Note that results from DINO and Yales 2 are also reported in the study. They are not included in the figures here for the sake of readability, but are in excellent agreement with NEK5000. 

\begin{figure*}[htbp]
    \centering
    \subfigure[]{
		\begin{minipage}[t]{0.3\linewidth}
			\centering
			\includegraphics[width=\linewidth]{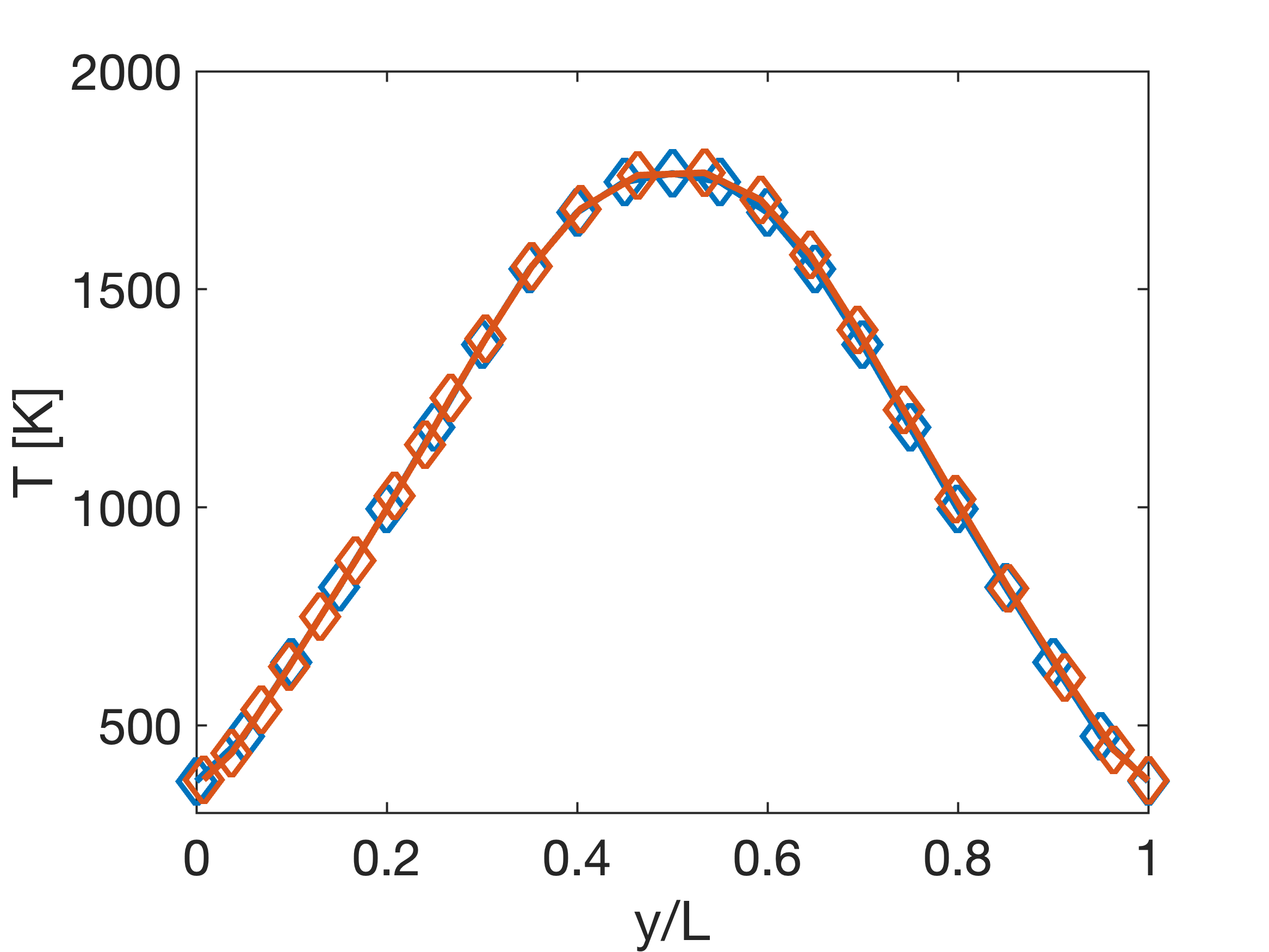}
		\end{minipage}
    }
    \subfigure[]{
		\begin{minipage}[t]{0.3\linewidth}
			\centering
			\includegraphics[width=\linewidth]{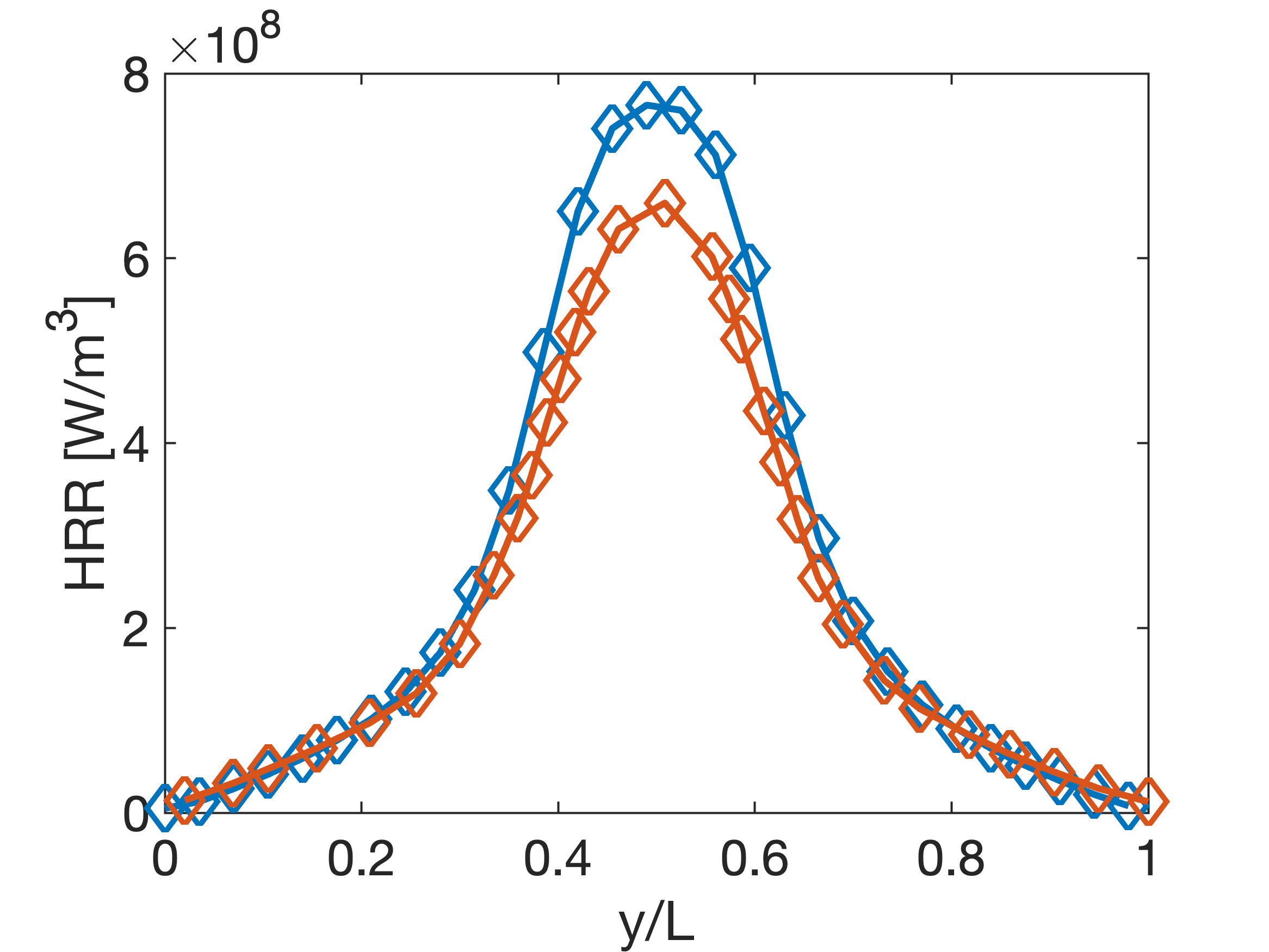}
		\end{minipage}
    }
    \subfigure[]{
		\begin{minipage}[t]{0.3\linewidth}
			\centering
			\includegraphics[width=\linewidth]{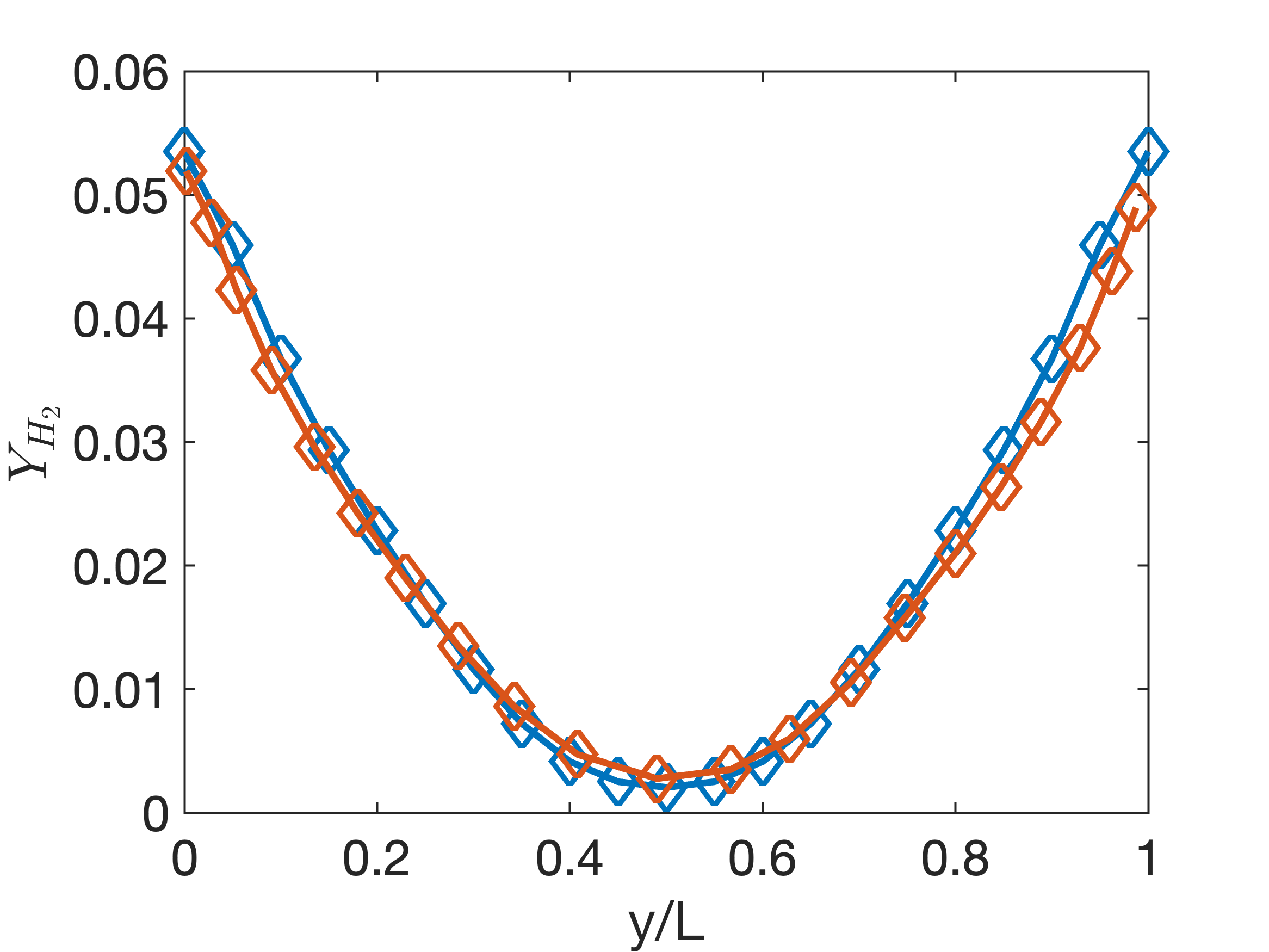}
		\end{minipage}
    }
        \subfigure[]{
		\begin{minipage}[t]{0.3\linewidth}
			\centering
			\includegraphics[width=\linewidth]{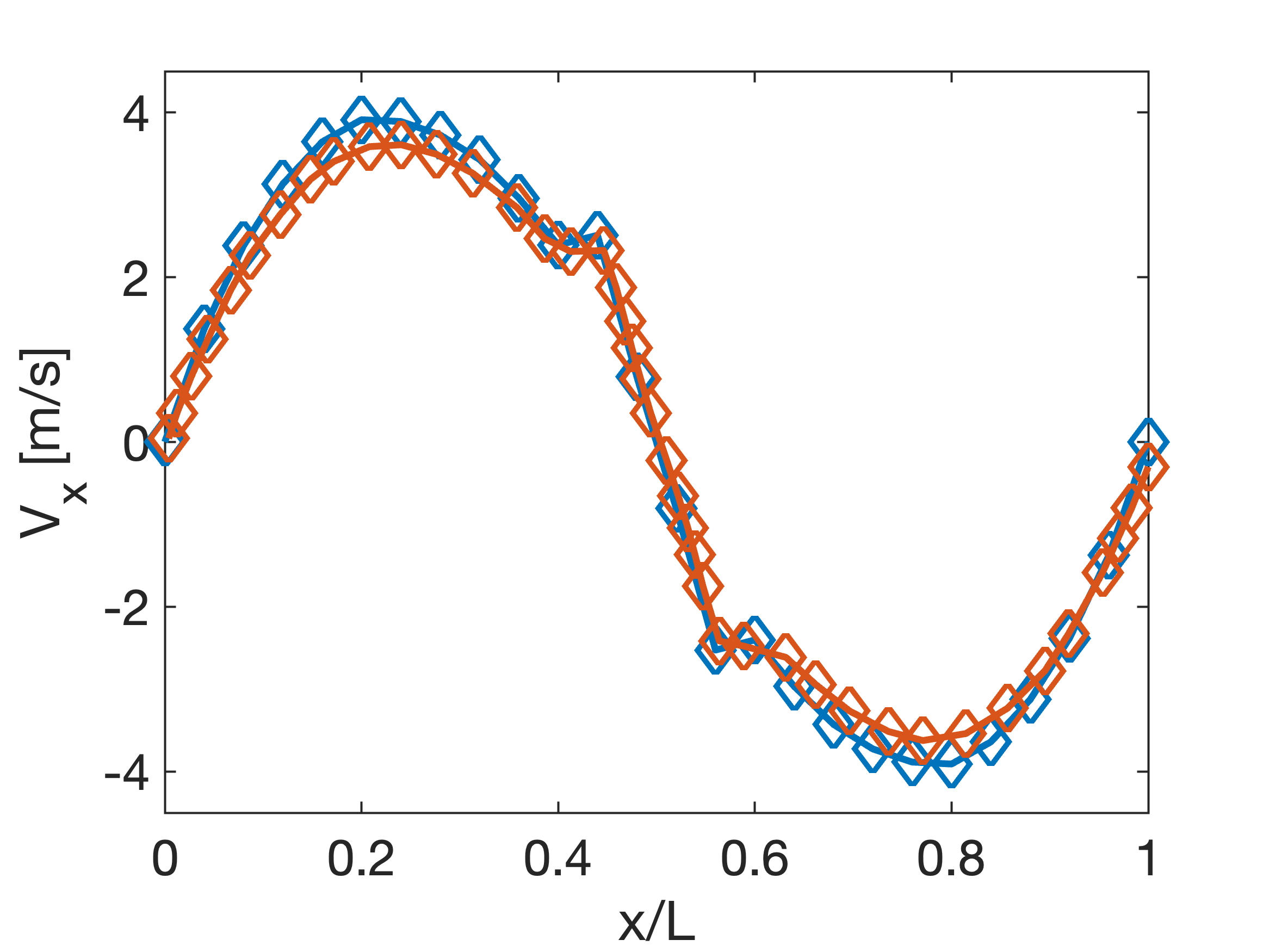}
		\end{minipage}
    }
    \subfigure[]{
		\begin{minipage}[t]{0.3\linewidth}
			\centering
			\includegraphics[width=\linewidth]{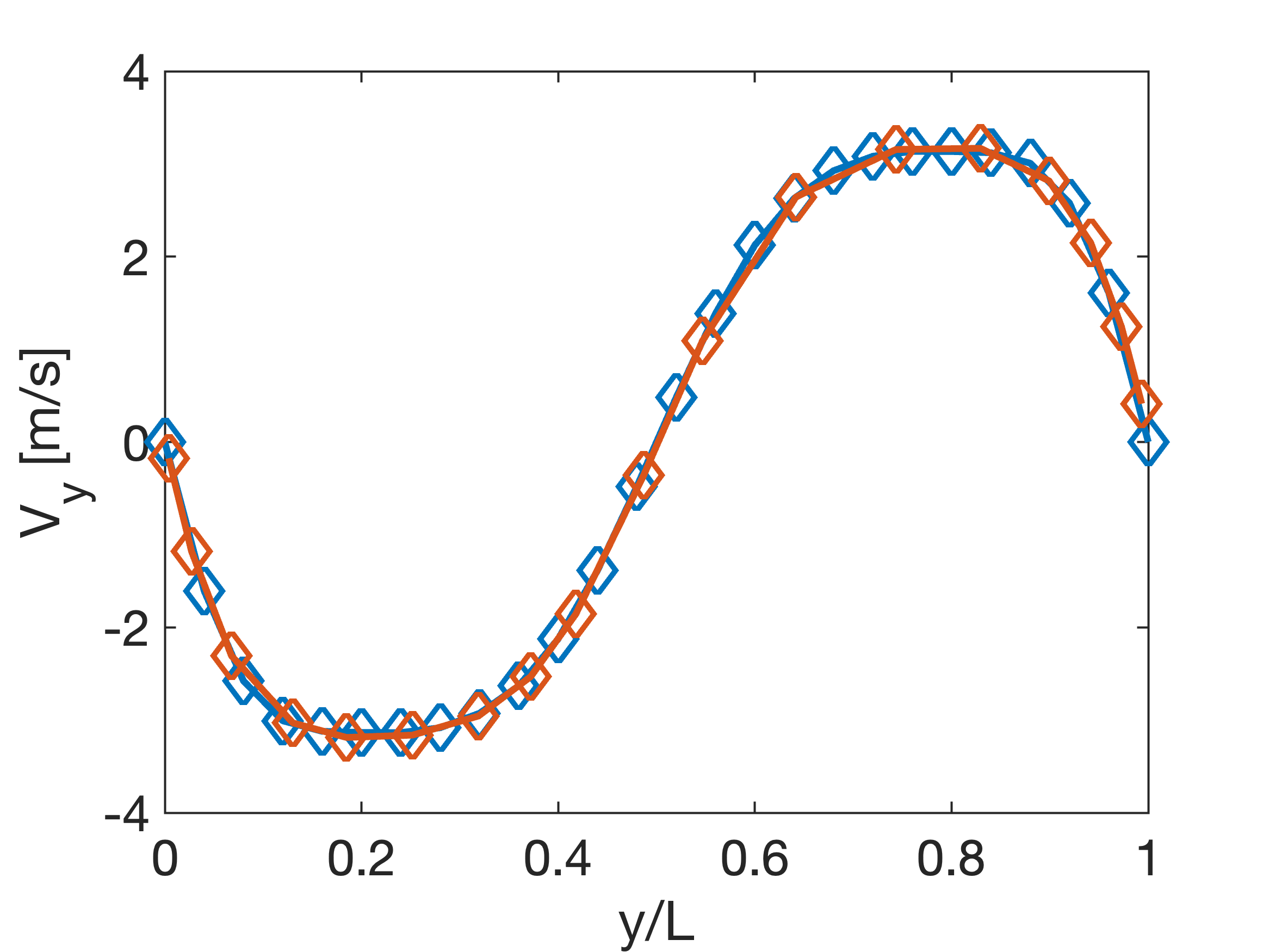}
		\end{minipage}
    }
    \centering
	\caption{Results of the reactive case at 0.5 ms. Profiles obtained on the middle axis identified in the bottom right of each plot: Temperature [K], heat release rate (HRR) [W/m$^3$], $Y_{H_2}$, $V_x$~[m/s] and $V_y$~[m/s]. ASTR (blue marked lines), NEK (red marked lines).}
	\label{fig14}
\end{figure*}

\begin{figure}[htbp]
\centering
\includegraphics[width=0.7\linewidth]{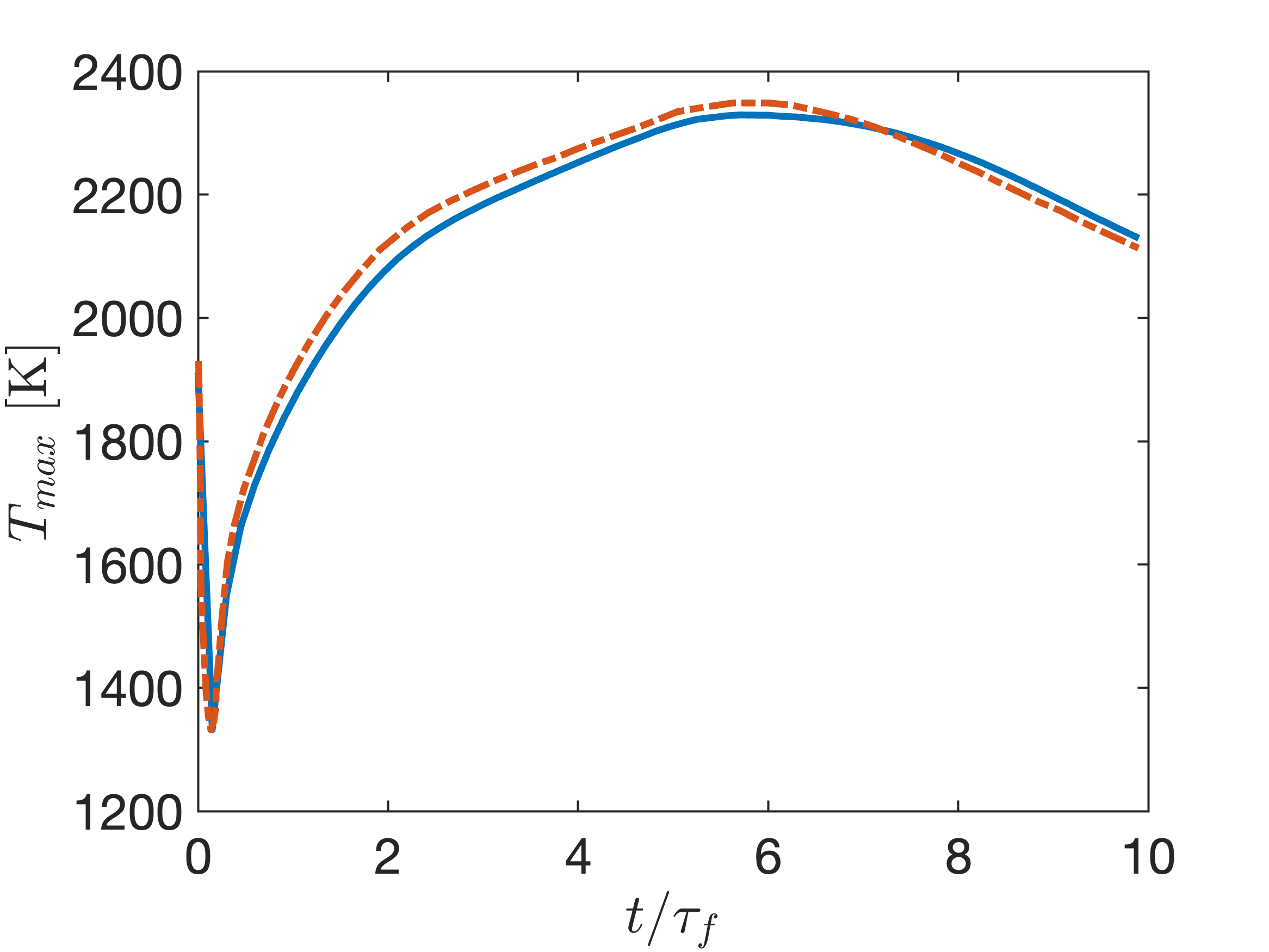}
\caption{Domain maximum temperature evolution over time. ASTR (blue solid line), NEK (red dot-dashed line).}
\label{fig15}
\end{figure}

Our simulations are performed with a constant time step of $\Delta t = 1.25\times10^{-8}$~s. Comparisons for the reacting multi-species flow are provided in Fig. \ref{fig14}, for $T$, heat release rate (HRR), $Y_{H_2}$, $V_x$ and $V_y$
centerline profiles. Figure \ref{fig15} shows the domain maximum temperature evolution over time. The results obtained by ASTR are marked in blue lines and the results obtained by NEK5000 reported are marked in red lines. According to Fig. \ref{fig14}, velocity profiles and temperature along the centerline show excellent agreement. Minor discrepancies are observed on $Y_{H_2}$ and heat release plots. Similar discrepancies can be observed in Fig. \ref{fig15}, which are seen to be within the code to code deviation for this resolution level.

\section*{Acknowledgement}
This work is supported by the National Science Foundation of China (Grant No. 52276096 and No. 92270203) and the Emerging Interdisciplinary-Young Scholars Project, Peking University, the Fundamental Research Funds for the Central Universities. Part of the numerical simulations was performed on the High Performance Computing Platform of CAPT, Peking University.

\bibliographystyle{ieeetr}
\bibliography{ref}

\begin{thebibliography}{10}

\bibitem{renard_dynamics_2000}
P.-H. Renard, D.~Thévenin, J.~Rolon, and S.~Candel, ``Dynamics of flame/vortex
  interactions,'' vol.~26, no.~3, pp.~225--282.

\bibitem{cuenot_effects_1994}
B.~Cuenot and T.~Poinsot, ``Effects of curvature and unsteadiness in diffusion
  flames. implications for turbulent diffusion combustion,'' vol.~25, no.~1,
  pp.~1383--1390.

\bibitem{kazbekov_enstrophy_2019}
A.~Kazbekov, K.~Kumashiro, and A.~M. Steinberg, ``Enstrophy transport in swirl
  combustion,'' vol.~876, pp.~715--732.

\bibitem{fillo_assessing_nodate}
A.~J. Fillo, P.~E. Hamlington, and K.~E. Niemeyer, ``Assessing diffusion model
  impacts on enstrophy and flame structure in turbulent lean premixed flames,''
  p.~17.

\bibitem{poinsot_quenching_1991}
T.~Poinsot, D.~Veynante, and S.~Candel, ``Quenching processes and premixed
  turbulent combustion diagrams,'' vol.~228, p.~561.

\bibitem{roberts_images_1993}
W.~Roberts, J.~Driscoll, M.~Drake, and L.~Goss, ``Images of the quenching of a
  flame by a vortex—to quantify regimes of turbulent combustion,'' vol.~94,
  no.~1, pp.~58--69.

\bibitem{hermanns_dynamics_2007}
M.~Hermanns, M.~Vera, and A.~Liñán, ``On the dynamics of flame edges in
  diffusion-flame/vortex interactions,'' vol.~149, no.~1, pp.~32--48.

\bibitem{AGOSTINELLI2022112120}
P.~Agostinelli, D.~Laera, I.~Chterev, I.~Boxx, L.~Gicquel, and T.~Poinsot, ``On
  the impact of h2-enrichment on flame structure and combustion dynamics of a
  lean partially-premixed turbulent swirling flame,'' {\em Combustion and
  Flame}, vol.~241, p.~112120, 2022.

\bibitem{ANIELLO2023112595}
A.~Aniello, D.~Laera, S.~Marragou, H.~Magnes, L.~Selle, T.~Schuller, and
  T.~Poinsot, ``Experimental and numerical investigation of two flame
  stabilization regimes observed in a dual swirl h2-air coaxial injector,''
  {\em Combustion and Flame}, vol.~249, p.~112595, 2023.

\bibitem{poinsot_theoretical_2005}
T.~Poinsot and D.~Veynante, {\em Theoretical and numerical combustion}.
\newblock RT Edwards, Inc., 2005.

\bibitem{DOMINGO2022}
P.~Domingo and L.~Vervisch, ``Recent developments in dns of turbulent
  combustion,'' {\em Proceedings of the Combustion Institute}, 2022.

\bibitem{TaylorGreen}
G.~I. Taylor and A.~E. Green, ``Mechanism of the production of small eddies
  from large ones,'' {\em Proceedings of the Royal Society of London. Series A
  - Mathematical and Physical Sciences}, vol.~158, no.~895, pp.~499--521, 1937.

\bibitem{WangTGV}
Z.~Wang, K.~Fidkowski, R.~Abgrall, F.~Bassi, D.~Caraeni, A.~Cary, H.~Deconinck,
  R.~Hartmann, K.~Hillewaert, H.~Huynh, N.~Kroll, G.~May, P.-O. Persson, B.~van
  Leer, and M.~Visbal, ``High-order cfd methods: current status and
  perspective,'' {\em International Journal for Numerical Methods in Fluids},
  vol.~72, no.~8, pp.~811--845, 2013.

\bibitem{sharma_vorticity_2019}
N.~Sharma and T.~K. Sengupta, ``Vorticity dynamics of the three-dimensional
  taylor-green vortex problem,'' vol.~31, no.~3, p.~035106.

\bibitem{brachet_taylor-green_1984}
M.~E. Brachet, D.~Meiron, S.~Orszag, B.~Nickel, R.~Morf, and U.~Frisch, ``The
  taylor-green vortex and fully developed turbulence,'' vol.~34, no.~5,
  pp.~1049--1063.

\bibitem{lusher_assessment_2021}
D.~J. Lusher and N.~D. Sandham, ``Assessment of low-dissipative shock-capturing
  schemes for the compressible taylor–green vortex,'' vol.~59, no.~2,
  pp.~533--545.

\bibitem{peng_effects_2018}
N.~Peng and Y.~Yang, ``Effects of the mach number on the evolution of
  vortex-surface fields in compressible taylor-green flows,'' vol.~3, no.~1,
  p.~013401.

\bibitem{yang_evolution_2011}
Y.~Yang and D.~I. Pullin, ``Evolution of vortex-surface fields in viscous
  taylor–green and kida–pelz flows,'' vol.~685, pp.~146--164.

\bibitem{abdelsamie_taylorgreen_2021}
A.~Abdelsamie, G.~Lartigue, C.~E. Frouzakis, and D.~Thévenin, ``The
  taylor–green vortex as a benchmark for high-fidelity combustion simulations
  using low-mach solvers,'' vol.~223, p.~104935.

\bibitem{mao_deepflame_2022}
R.~Mao, M.~Lin, Y.~Zhang, T.~Zhang, Z.-Q.~J. Xu, and Z.~X. Chen, ``{DeepFlame}:
  {A} deep learning empowered open-source platform for reacting flow
  simulations,'' Oct. 2022.
\newblock arXiv:2210.07094 [physics].

\bibitem{boivin2021benchmarking}
P.~Boivin, M.~Tayyab, and S.~Zhao, ``Benchmarking a lattice-boltzmann solver
  for reactive flows: Is the method worth the effort for combustion?,'' {\em
  Physics of Fluids}, vol.~33, no.~7, p.~071703, 2021.

\bibitem{cuenot_asymptotic_nodate}
B.~Cuenot and T.~Poinsot, ``Asymptotic and numerical study of diffusion flames
  with variable lewis number and finite rate chemistry,'' p.~27.

\bibitem{han_thermal_2021}
W.~Han, A.~Scholtissek, F.~Dietzsch, and C.~Hasse, ``Thermal and chemical
  effects of differential diffusion in turbulent non-premixed {H2} flames,''
  {\em Proceedings of the Combustion Institute}, vol.~38, no.~2,
  pp.~2627--2634, 2021.

\bibitem{domingo_recent_nodate}
P.~Domingo, ``Recent developments in {DNS} of turbulent combustion,'' p.~22.

\bibitem{poinsot_applications_nodate}
T.~Poinsot, S.~Candelt, and A.~Trouvt, ``{APPLICATIONS} {OF} {DIRECT}
  {NUMERICAL} {SIMULATION} {TO} {PREMIXED} {TURBULENT} {COMBUSTION},'' p.~46.

\bibitem{ZHANG2022112319}
T.~Zhang, Y.~Yi, Y.~Xu, Z.~X. Chen, Y.~Zhang, W.~E, and Z.-Q.~J. Xu, ``A
  multi-scale sampling method for accurate and robust deep neural network to
  predict combustion chemical kinetics,'' {\em Combustion and Flame}, vol.~245,
  p.~112319, 2022.

\bibitem{fang_improved_2019}
J.~Fang, F.~Gao, C.~Moulinec, and D.~Emerson, ``An improved parallel compact
  scheme for domain‐decoupled simulation of turbulence,'' vol.~90, no.~10,
  pp.~479--500.

\bibitem{fang_optimized_2013}
J.~Fang, Z.~Li, and L.~Lu, ``An optimized low-dissipation
  monotonicity-preserving scheme for numerical simulations of high-speed
  turbulent flows,'' vol.~56, no.~1, pp.~67--95.

\bibitem{fang_turbulence_2020}
J.~Fang, A.~A. Zheltovodov, Y.~Yao, C.~Moulinec, and D.~R. Emerson, ``On the
  turbulence amplification in shock-wave/turbulent boundary layer
  interaction,'' vol.~897, p.~A32.

\bibitem{BOIVIN2011517}
P.~Boivin, C.~Jiménez, A.~Sánchez, and F.~Williams, ``An explicit reduced
  mechanism for h2–air combustion,'' {\em Proceedings of the Combustion
  Institute}, vol.~33, no.~1, pp.~517--523, 2011.

\end{thebibliography}

\end{document}